\documentclass[aps,showpacs,pre,amsmath,twocolumn,amssymb,floatfix,superscriptaddress]{revtex4}

\usepackage{epsfig}
\usepackage{graphics}
\usepackage{graphicx,psfrag}
\usepackage{dcolumn}
\usepackage{bm}

\newcommand{\beq}{\begin{equation}}
\newcommand{\eeq}{\end{equation}}
\newcommand{\bea}{\begin{eqnarray}}
\newcommand{\eea}{\end{eqnarray}}

\newcommand{\nn}{\nonumber\\}

\begin{document}

\title{On explosion of the chaotic attractor}

\author{P. Badank\'o and K. Sailer}

\affiliation{Department of Theoretical Physics, University of Debrecen,
Debrecen, Hungary}

\date{\today}

\begin{abstract}
There are presented examples of the 
rather sudden and violent explosion of the strange attractor
 of a one-dimensional  driven damped  anharmonic  oscillator
 induced by a relatively small 
change of the amplitude of the strongly nonperturbative periodic driving force.
A phenomenologic characterization of the explosion of the strange attractor has been 
given in terms  of the behavior of the average maximal Lyapunov exponent ${\bar \lambda}$
 and that of the fractal dimension $D_{q}$ for $q=-4$.    It is 
shown that the building up of the exploding strange attractor is accompanied by a  nearly
 linear increase  of the  maximal average Lyapunov exponent ${\bar \lambda}$. A sudden jump 
of the fractal dimension $D_{-4}$ is detected  when the explosion starts off from
 an attractor consisting of disjoint bunches separated by an empty phase-space region.

\end{abstract}

\pacs{05.45.Ac} 

\maketitle

\section{Introduction}

Various cases of the one-dimensional periodically driven damped anharmonic  oscillator like
 the Duffing and the van der Pol oscillators are
well-known and thoroughly investigated dissipative systems of relatively simple kinematics
 in which deterministic chaos appears for particular values of the parameters 
\cite{Ueda65,Haya70, Ueda73,Ueda79,Ueda80,Nova82,Test82,Freh85,Ueda85, Tomita86,Ueda91,
Laksh97,Vank03,Bona08}.
In the present paper we would like to concentrate our attention to
 the explosion of the strange attractor under which we mean  the following phenomenon.
Supposing that for a particular value of some control parameter the strange attractor  
consists of highly populated disjoint  bunches separated
by   an empty or underpopulated phase-space region and
 the latter gets  filled in gradually  with the continuous change of the control
 parameter. Our concept of the explosion of the strange attractor is here more qualitative
 and general than that introduced originally in \cite{Ueda80}. There a similar phenomenon
 has been reported for  particular values of the parameters of the damped, purely 
 quartic oscillator (called here oscillator O2)
under the simultaneous exertion of a periodic and a constant driving forces, the latter serving
 the control parameter. The relatively weak driving forces
 enabled the author to give a detailed dynamical explanation of the explosion of the
 strange attractor. Namely, it arises when the 
unstable and stable regions of a hyperbolic fixed point happen to touch tangentially 
with the increase of the control parameter. It has been found that the explosion of the
 strange attractor is interrupted time-to-time by windows of regular motion when the 
control parameter increases. 

 Here we show that apart from its dynamical origin a similar phenomenon occurs for particular 
parameter sets of the periodically 
driven damped  quartic oscillator  with a
nonvanishing  linear component of the restoring force, but with a vanishing constant external force 
 (called  oscillator O1 below). Now the amplitude of the
periodic driving force controls the explosion of the strange attractor and it
has extremely large values that the driving is far away from being a perturbation,
as opposed to  the case of explosion for  oscillator O2 in  \cite{Ueda80}.
For oscillator O1 we present  cases when the highly populated bunches
 of the unexploded strange attractor are  separated by an  empty phase-space region that
is filled in rather suddenly when the control parameter
is increased by a relatively small amount, much less than this happens in the case of the explosion of the strange attractor presented in \cite{Ueda80}.
 Our term {\em empty phase-space region} needs some explanation. In the cases of exploding strange attractors belonging to the oscillator O1 we have established that the highly populated bunches of the yet unexploded  attractors are separated by a   phase-space region avoided  by the trajectory points. 
 For  numerical samplings with similar statistics only a severely underpopulated
region of the disjoint bunches has been found for the unexploded strange attractor
for oscillator O2 given in  \cite{Ueda80}.
 Therefore, we shall say that
in the examples found by us for oscillator O1 the explosion starts from an unexploded
 strange attractor
characterized by an {\em empty} phase-space region separating the highly populated bunches.  
Moreover, in the examples
 that we are going to present no windows of periodic motion interrupt the explosion.
We shall show that in these examples the explosion is more sudden and violent according to its
 quantitative characteristics as compared to the example discussed in  \cite{Ueda80}.

 The qualitatively similar features of the explosion of the strange attractor found by us
 for the cases of  oscillator O1 may or may not be the consequence of dynamics similar to
that for oscillator O2.
 In our  examples a detailed dynamical analysis of the explosion seems to be hardly available
because of the  strongly nonperturbative nature of the driving 
 as opposed to the case discussed  in \cite{Ueda80}.
 Here our purpose is to show that there exist  rather phenomenologic static global
 characteristics of the exploding strange  attractor which are sensitive to the
gradual change of the control parameter. We are  showing that such characteristics are the
 average maximal Lyapunov exponent, as well as the  generalized dimension $D_q$ for
some asymptotically large negative $q$ value, (at $q=-4$ in our work). 
For a comparison, we analyse the explosion of the strange attractor for both types of oscillators
 O1 and O2. 

Our paper is constructed as follows. In Sect. \ref{oscis} we describe the externally driven,
damped, anharmonic oscillators under consideration and the numerics used for the solution of 
their equations of motion. Also the qualitative features of the explosion of the strange
 attractor
are discussed for the various cases. Sect. \ref{Lyap} presents the details of the 
determination of the maximal Lyapunov exponents averaged over various initial conditions
and its tendency of monotonic increase during the building up of the exploding strange
 attractor is shown. In Sect. \ref{gendim} the details of the numerical  determination of
 the fractal dimensions $D_q$ are given and a sudden jump of the fractal dimension $D_q$ 
with negative quotient $q=-4$ is shown for cases when the explosion starts off from
 an attractor of disjoint bunches separated by an empty phase-space region. A comparison
of the various cases of the explosions of the strange attractor is given on the scale
of the relative change of the control parameter. Finally, the conclusions are drawn in
Sect. \ref{conclu}.

\section{Exploding strange attractors for Duffing oscillators}\label{oscis}

Both oscillators O1 and O2 investigated by us are those of Duffing-type with a single-well 
potential. In terms of dimensionless parameters the equations of motion 
 of oscillator O1 are
\bea
  {\dot x}&=&v,\nn
  {\dot v}&=&-\gamma v - \delta^2 (x+x^3) + I\cos (2\pi t).
\eea
with the damping factor $\gamma$, the amplitude $I$ of
 the periodic external driving  force  and the parameter $\delta^2$ of the anharmonic
 potential. 
Here the initial conditions $x(0)=v(0)=1.0$ at $t=0$ were chosen to get an overall picture
 on the position and extension of  the bounded phase-space region in which the trajectory
 runs. Chaotic behaviour has been found by us for two sets of the parameters: {\em (A)} 
$\gamma=0.325,~\delta=3.25$ and the control parameter $I\in \lbrack 787.50, 788.75 \rbrack$ increased 
with the steps $\Delta I\approx 0.015 $, and {\em (B)}
$\gamma=0.65,~\delta=6.50$ and $I\in\lbrack 1924.0,1940.0 \rbrack$ varied with the steps 
$\Delta I =1.0 $.  No  windows of regular motion occurred in these intervals. 
The equations of motion of oscillator O2
 (see \cite{Ueda80}) are given as
\bea
  {\dot x} &=&v,\nn
 {\dot v}&=& -\gamma v-x^3 + I_0 +I \cos (2\pi t)
\eea
with  the amplitude $I_0$  of the constant 
driving  force, and the other parameters having the same meaning as for oscillator O1. We have chosen parameter set {\em (C)}  used in 
 \cite{Ueda80} with the fixed values $\gamma=0.05$ and $I=0.16$  
 and the control parameter $I_0$ variing in the interval $\lbrack 0.026, 0.055\rbrack$ with
 the discrete steps  $\Delta I_0\approx 0.001$.  The initial conditions $x(0)=v(0)=0.1$ at
 $t=0$ were chosen in the close neighbourhood of
 the chaotic attractor found in  \cite{Ueda80}.

\begin{center} 
\begin{figure*}[htb]
\epsfig{file=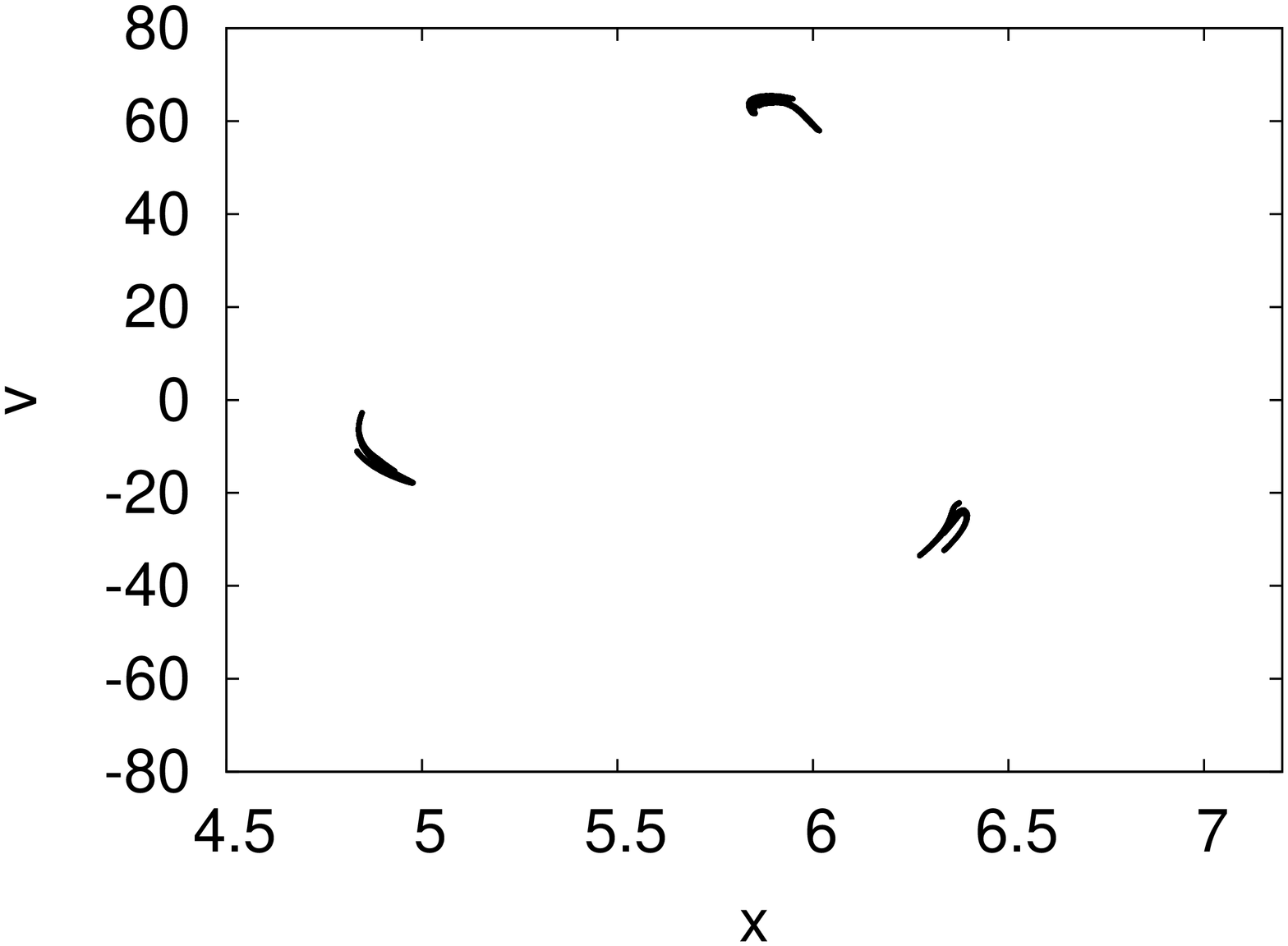,width=6cm,angle=-90}
\epsfig{file=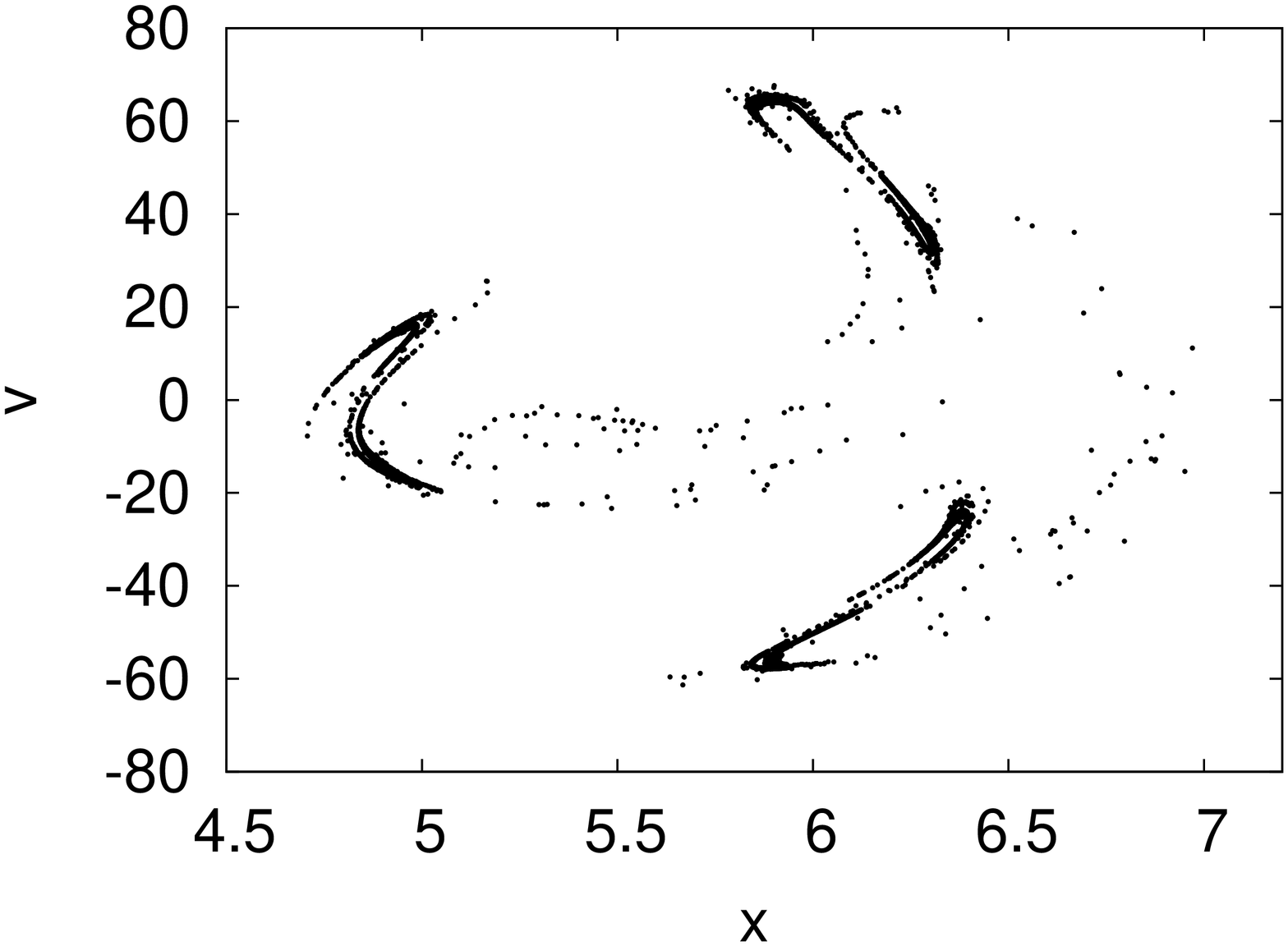,width=6cm,angle=-90}
\epsfig{file=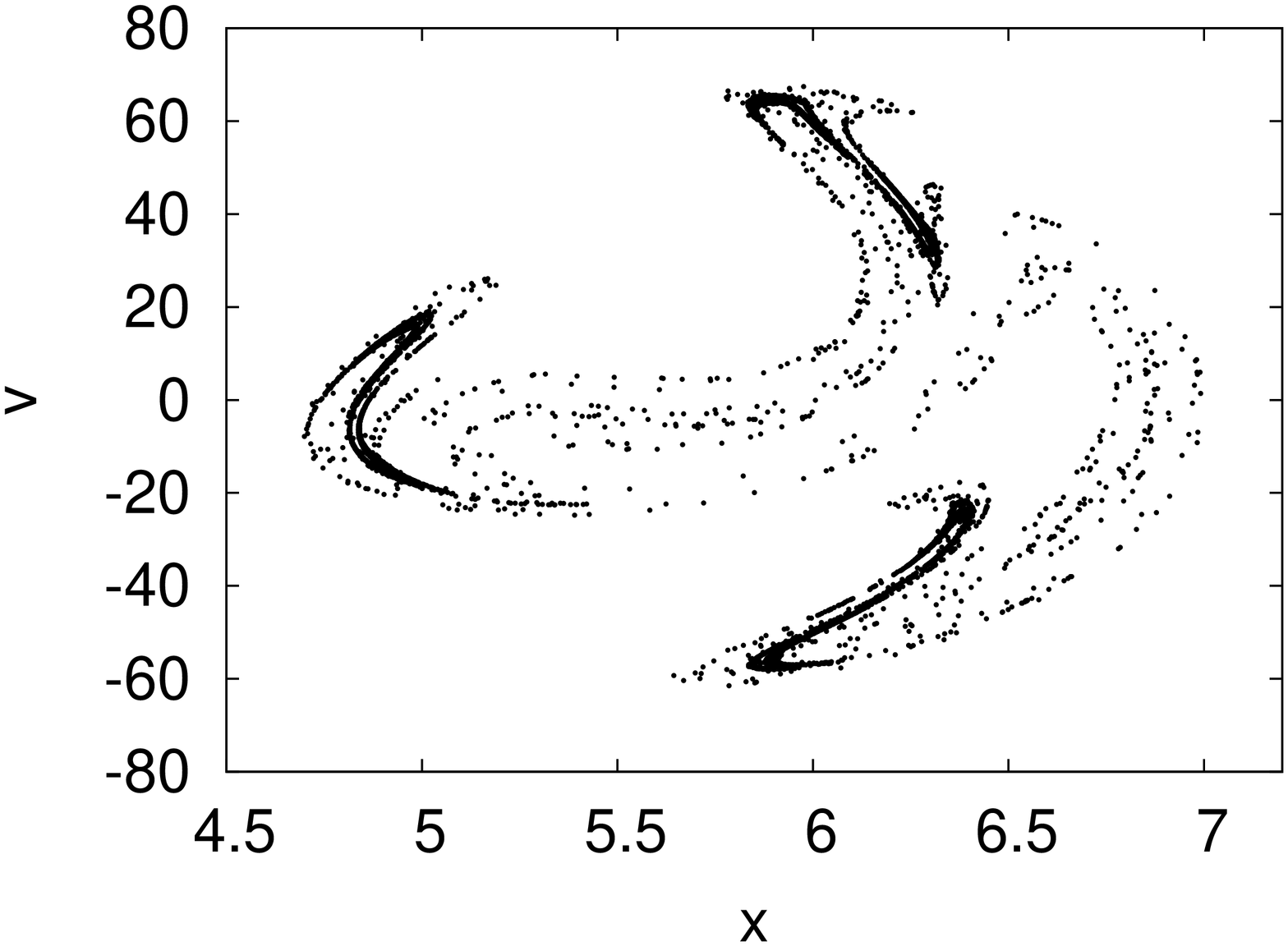,width=6cm,angle=-90}
\epsfig{file=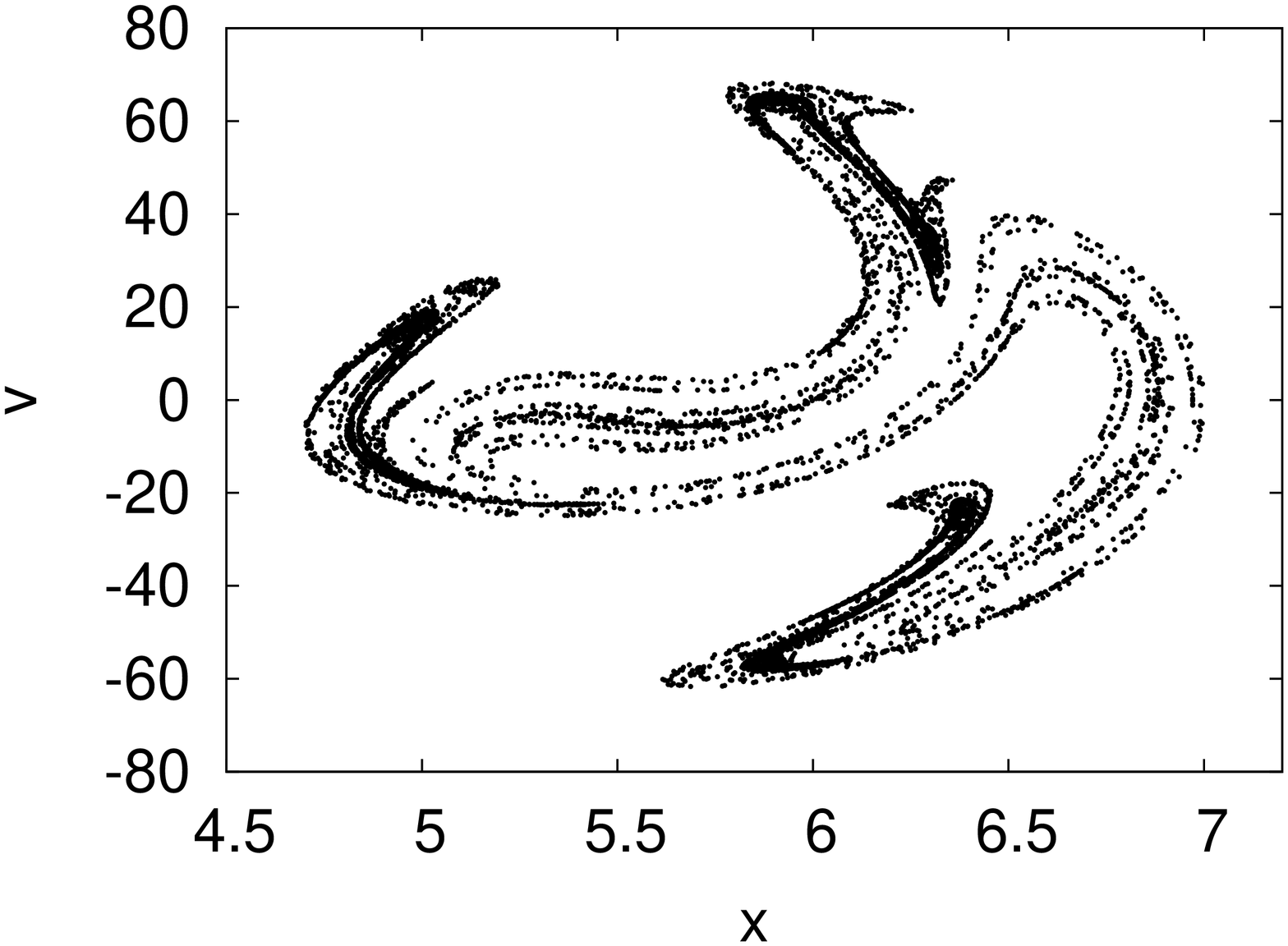,width=6cm,angle=-90}
\caption{\label{fig_attO1A} Explosion of the chaotic attractor for oscillator O1 with 
parameter set {\em (A)} and for the control-parameter values  $I= 787.91,~
788.01,~ 788.10,~ 788.400$
from the left to the right 
and from the top to the bottom, respectively.  }
\end{figure*}
\end{center}

The equations of motion have been numerically integrated by fourth-order Runge-Kutta 
method. The usage of the time step $\Delta t= 0.001$ provided stable regular solutions, 
as well as stable distribution of the points on the strange attractor. The time 
evolution of the phase trajectories has been followed through $\sim 10^4-10^5$ time periods of
 the periodic driving force 
after the initial transient has been damped at around $t_{damp}\approx 10^3$ time periods.
 The Poincar\'e-sections were determined by projecting the points of the phase trajectory
at each period of time, i.e., at times $t_n=t_0+ n+(\Delta \phi/2\pi)$ of the driving force onto 
the $(x,~v)$ plane, were the first Poincar\'e section has always been taken at the time 
$t_0 > t_{damp}$.
 The phase shift
was set to  $\Delta \phi=0$ except of the determination of the fractal dimension $D_{-4}$
when an average over various choices of $\Delta \phi$ has been taken (see below in
Sect. \ref{gendim} in more detail).
  The final decision on  the  regular or chaotic nature of the trajectory was
 taken depending on the sign of the maximal Lyapunov exponent (see the discussion in 
Sect. \ref{Lyap} below).

\begin{figure*}[htb]
\epsfig{file=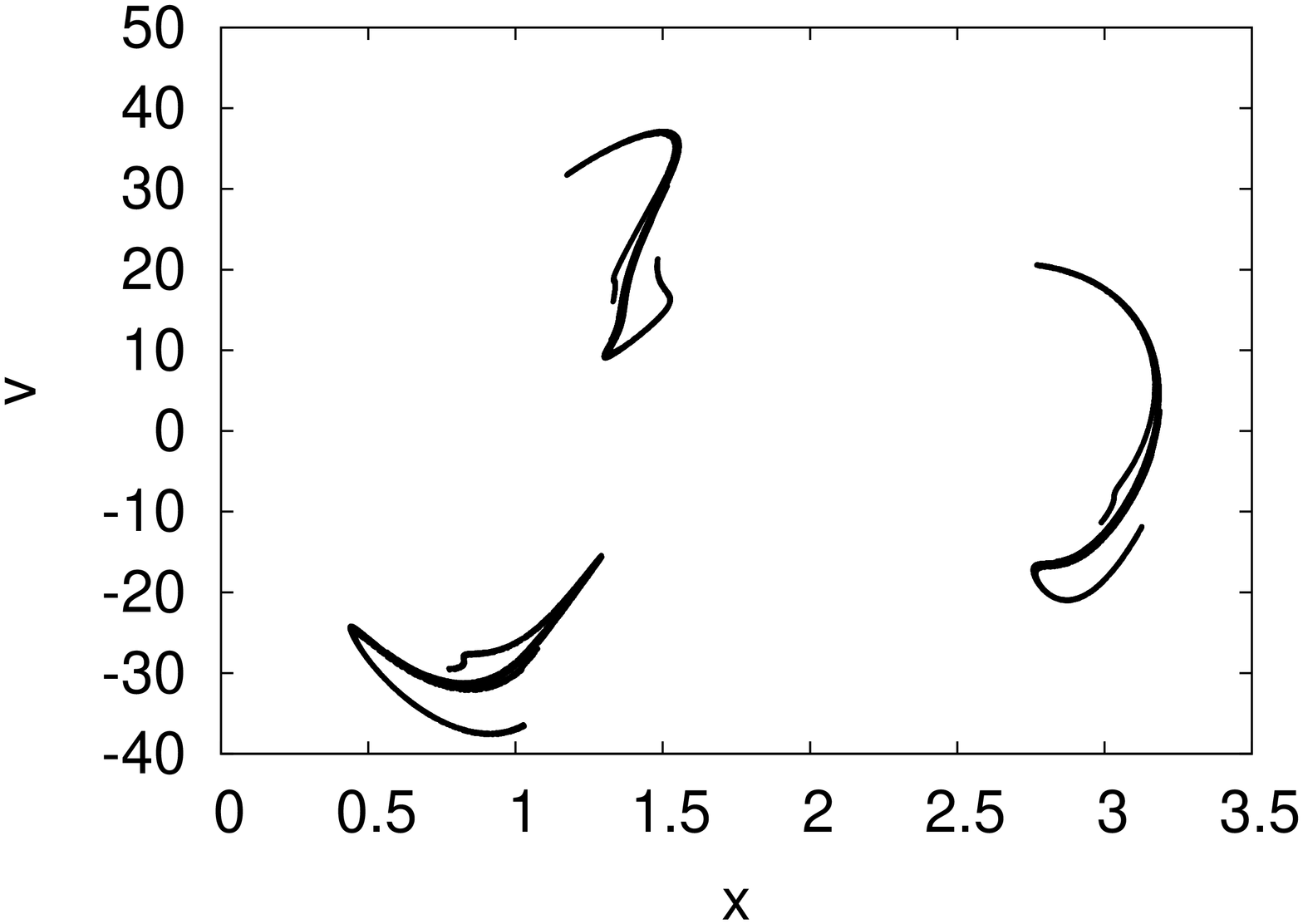,width=6cm,angle=-90}
\epsfig{file=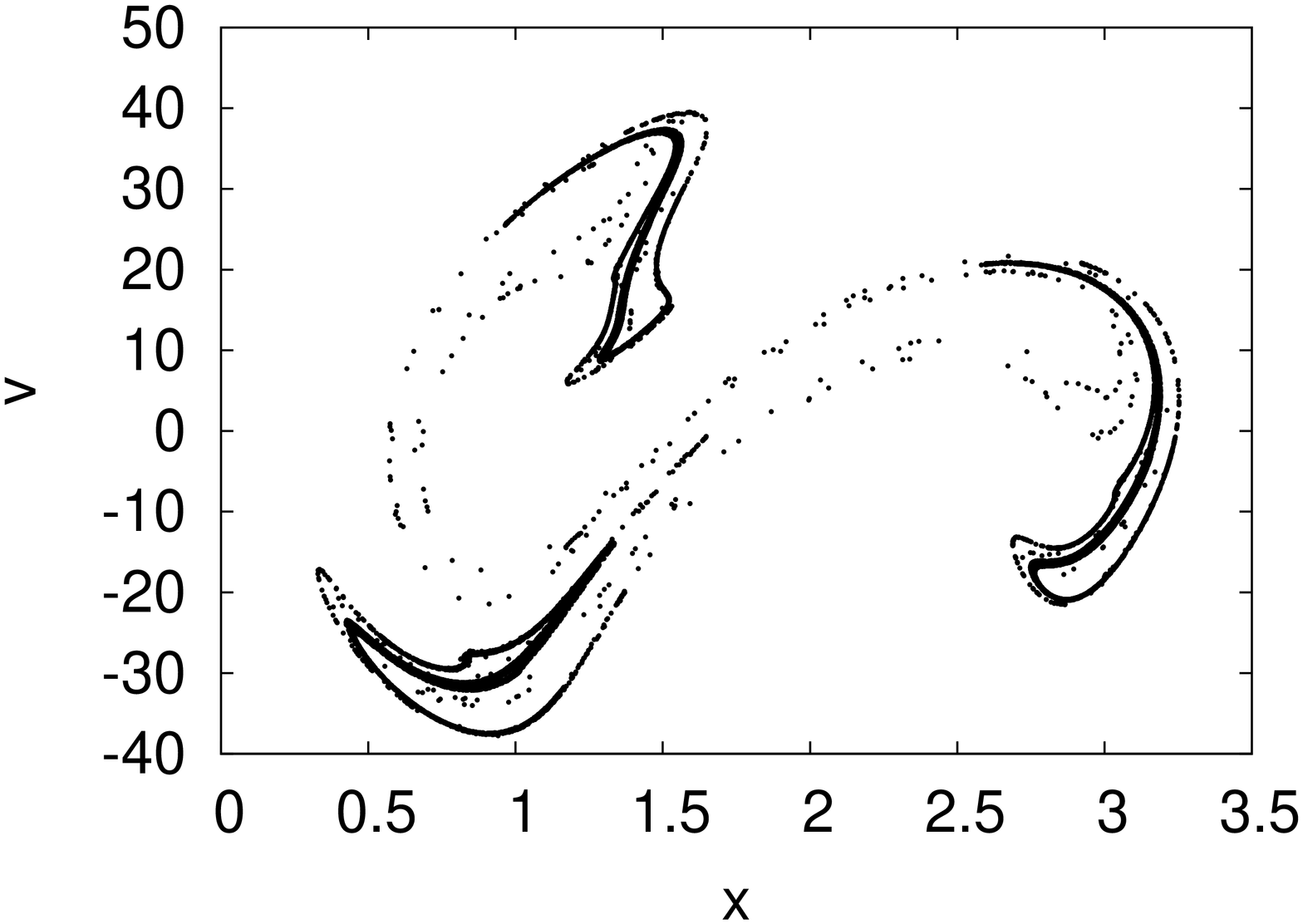,width=6cm,angle=-90}
\epsfig{file=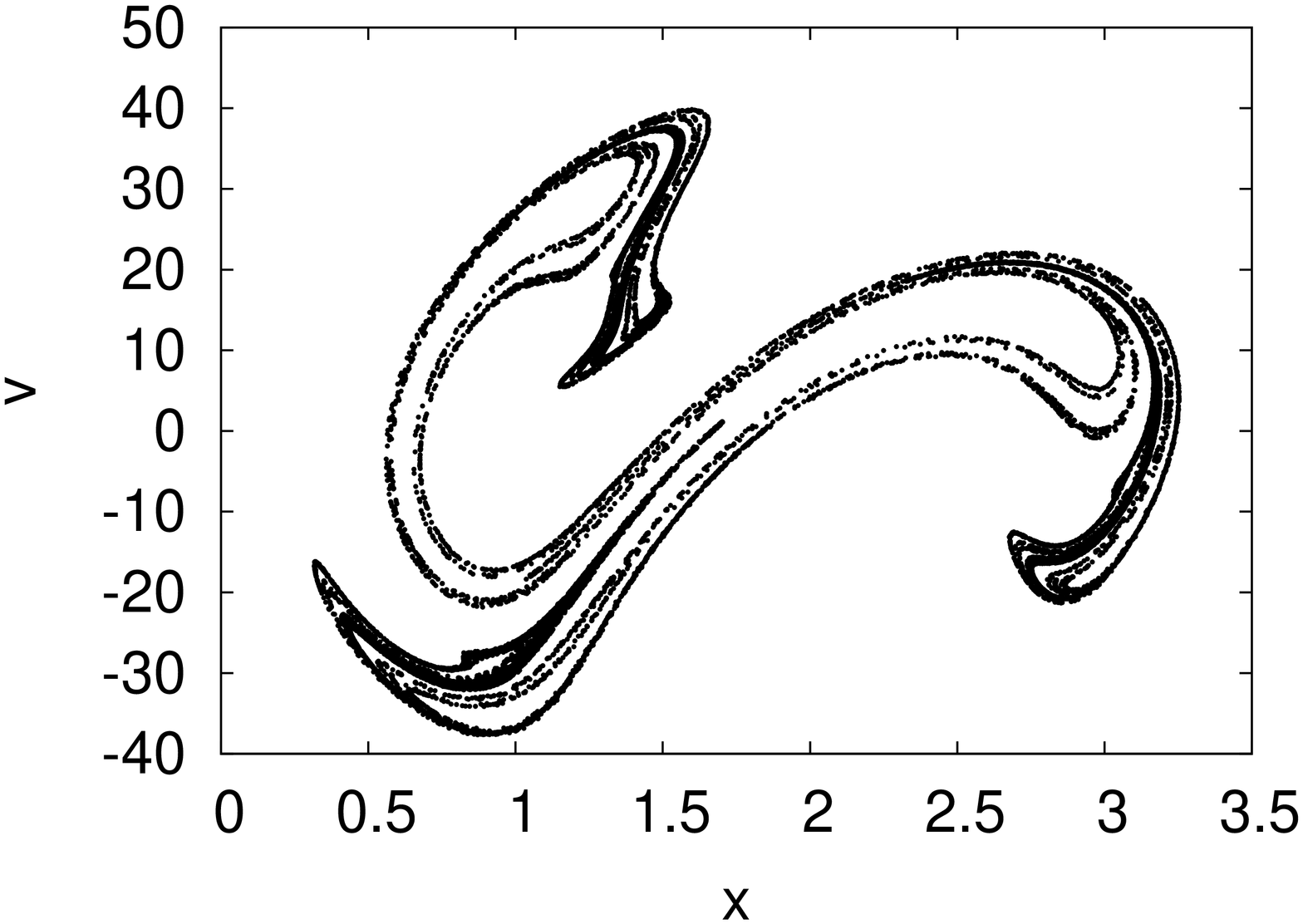,width=6cm,angle=-90}
\epsfig{file=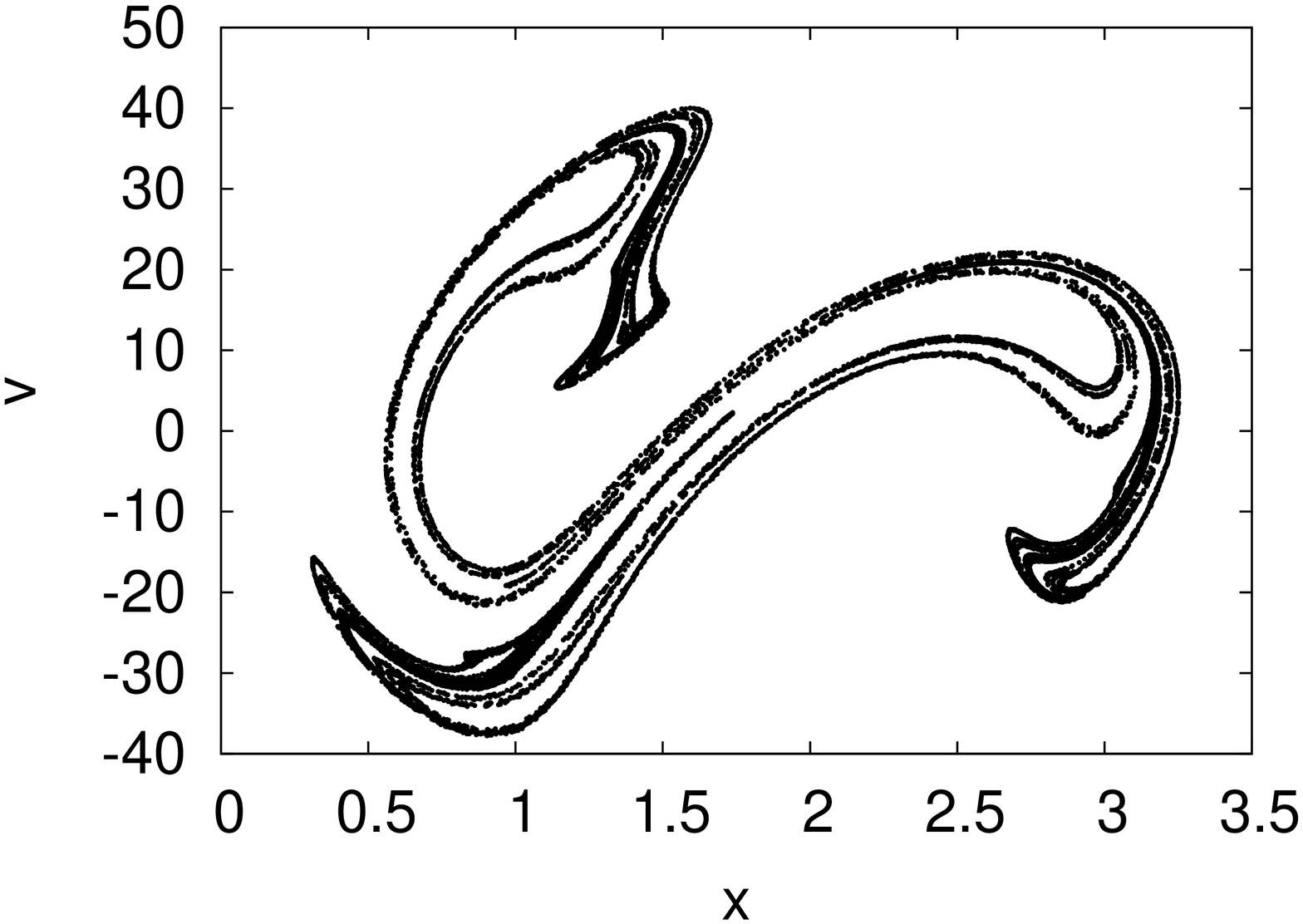,width=6cm,angle=-90}
\caption{\label{fig_attO1B} Explosion of the chaotic attractor for oscillator O1 with
parameter set {\em (B)} and for the control parameter values  $I=
 1929, 1930, 1933,~1935$
from the left to the right 
and from the top to the bottom,  respectively.}
\end{figure*}

Let us start with the description of the evolution of the strange attractor for oscillator 
O1 for parameter sets {\em (A)} and {\em (B)} illustrated in Figs. \ref{fig_attO1A} and
 \ref{fig_attO1B}. Below we 
shall determine the maximal Lyapunov exponent and establish its definitively positive value
 in all of the cases shown in these figures. For oscillator O1 and parameter set {\em (A)}
 the explosion  occurs with increasing control parameter $I$ as shown in Figs.
 \ref{fig_attO1A}. 
For $I=787.91$ the chaotic attractor is   initially concentrated   in  three 
 disjoint bunches on the Poincar\'e map separated by an essentially empty
phase-space region and with the increase of the control parameter from $I\approx 788.0$
 to $788.4$ it bursts out  filling in the whole phase-space region between the originally 
 disjoint bunches. It was also observed  that the strange attractor before the explosion
for $I\in \lbrack 787.50, 787.91\rbrack$ and just after it for
 $I\in \lbrack 788.4, 788.6\rbrack$ remains qualitatively unaltered. The critical value
is at $I_c\approx 788.0$ where the explosion of the attractor sets on and it continues
to $I\approx 788.4$.

For oscillator O1 with parameter set {\em (B)} a similar explosion of the strange attractor
occurs with increasing   control parameter $I$ (see Fig. \ref{fig_attO1B}). For values
 $I\in \lbrack 1926, 1929\rbrack$
 the strange attractor consists of three disjoint bunches, for $I$ raising from $1929$   
to $1935$  it bursts out, and
for $I \in \lbrack1935, 1938\rbrack$ its extension and picturial form are essentially kept.
The critical value is at $I_c\approx 1929.5$ where the explosion sets on and it lasts until
the value $I\approx 1935$ is reached. 

Now let us have  a closer look on the evolution of the strange attractor for the oscillator 
O2 with parameter set {\em (C)} used in \cite{Ueda80}. In this case  windows
 of regular and irregular motion alternate in the investigated parameter interval
 $I_0\in \lbrack  0.03, 0.045\rbrack$ which interrupt the explosion process
of the  strange attractor. The evolution of the attractor with the gradual increase of the
 control parameter $I_0$
 has been mapped
by the steps  $\Delta I_0=0.001$ and more densely  with the steps $\Delta I_0=0.00016$
in the intervals where the average maximal Lyapunov exponent ${\bar \lambda}$
 changes rather  rapidly.  Taking $I_0$ values from the regular windows, we have performed a
 thorough search for an additional strange attractor:
 we  have started several trajectories with various initial conditions distributed 
 in the phase-space region generally occupied by the strange attractor when it appears
with certainty for close values of the control parameter,
 but we have always found only regular trajectories. Therefore it can be excluded
that in these windows both a limit cycle and a strange attractor exist, as far as  the
investigated phase-space region is considered.

\begin{center} 
\begin{figure*}[htb]
\epsfig{file=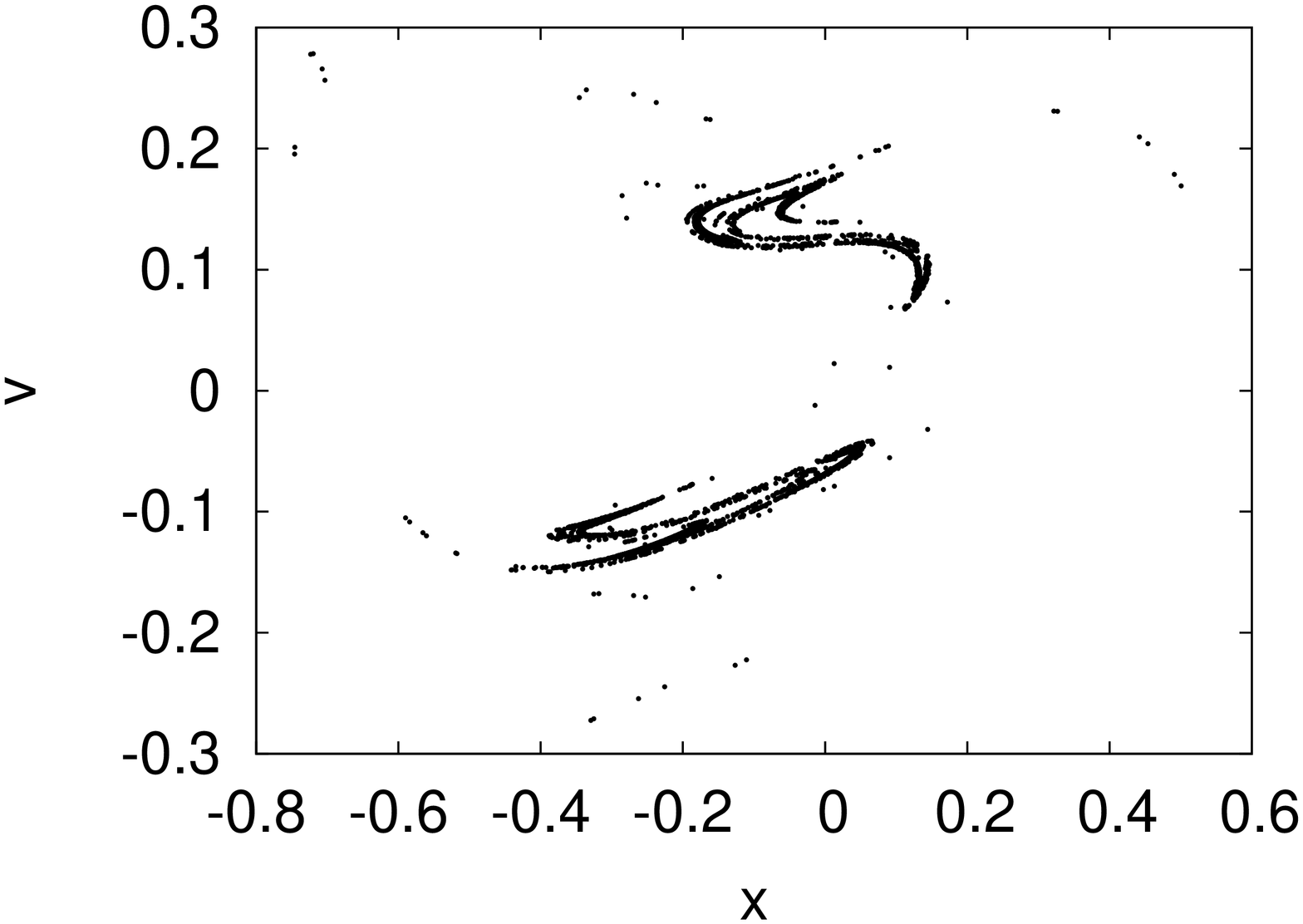,width=6cm,angle=-90}
\epsfig{file=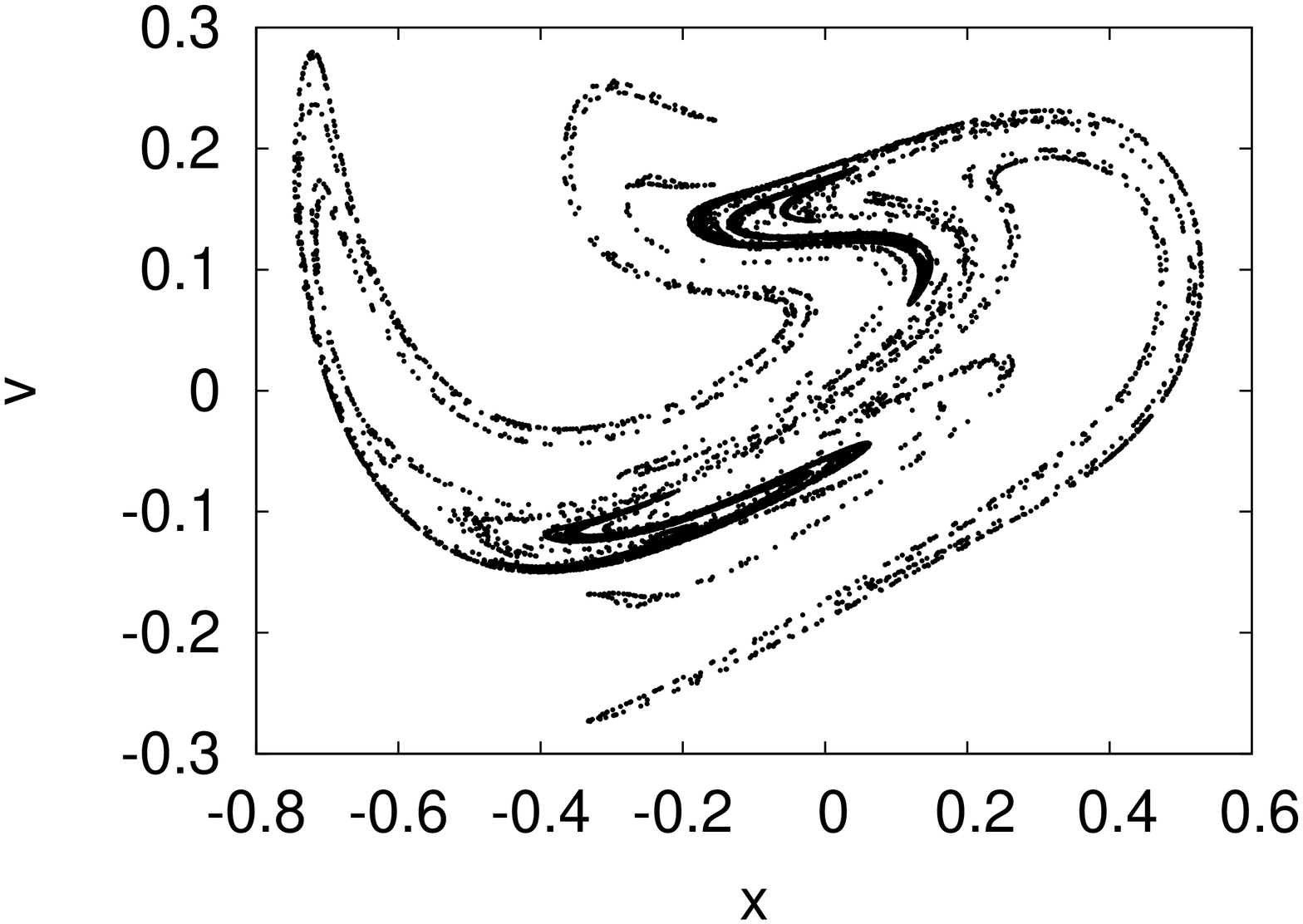,width=6cm,angle=-90}
\epsfig{file=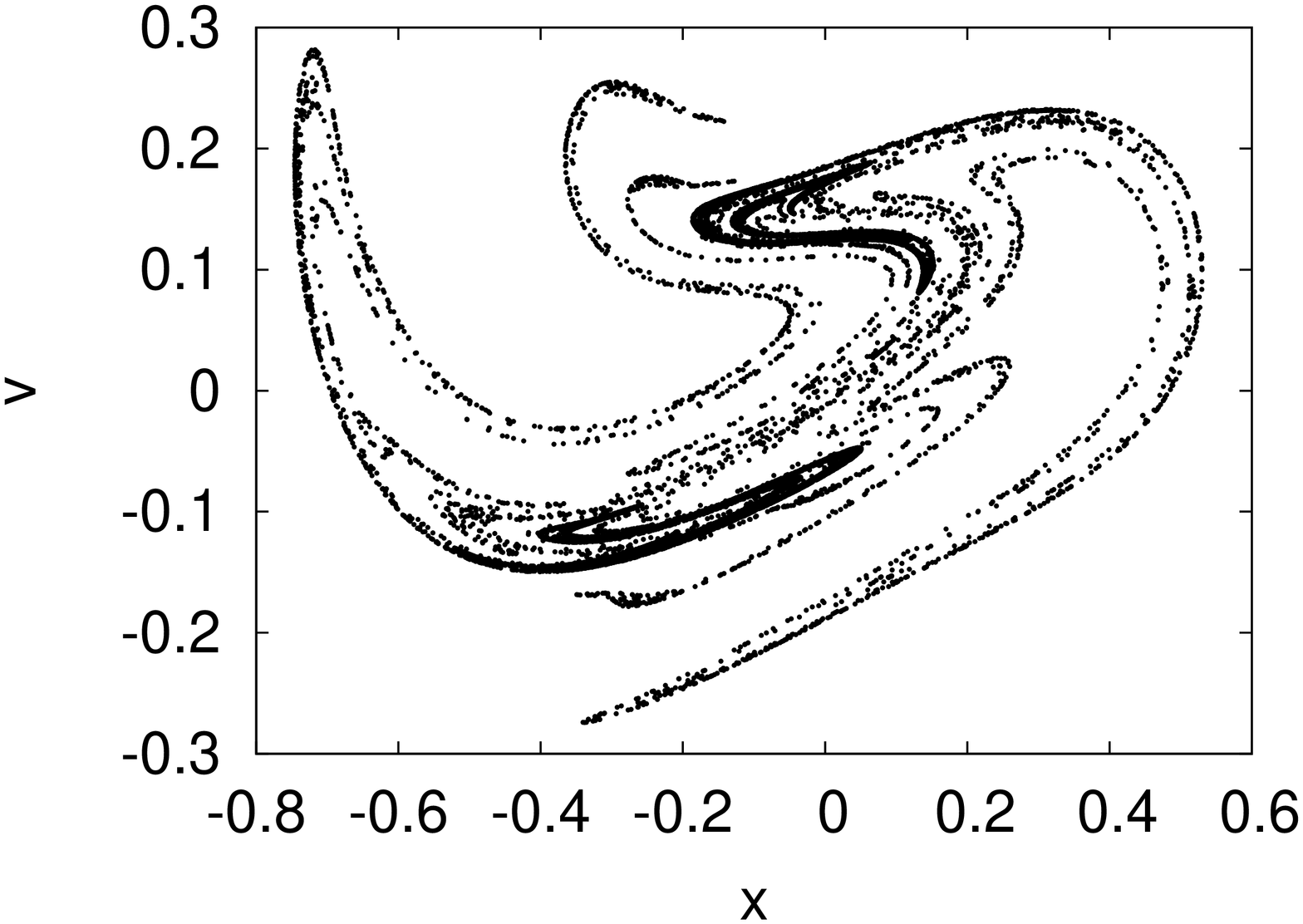,width=6cm,angle=-90}
\epsfig{file=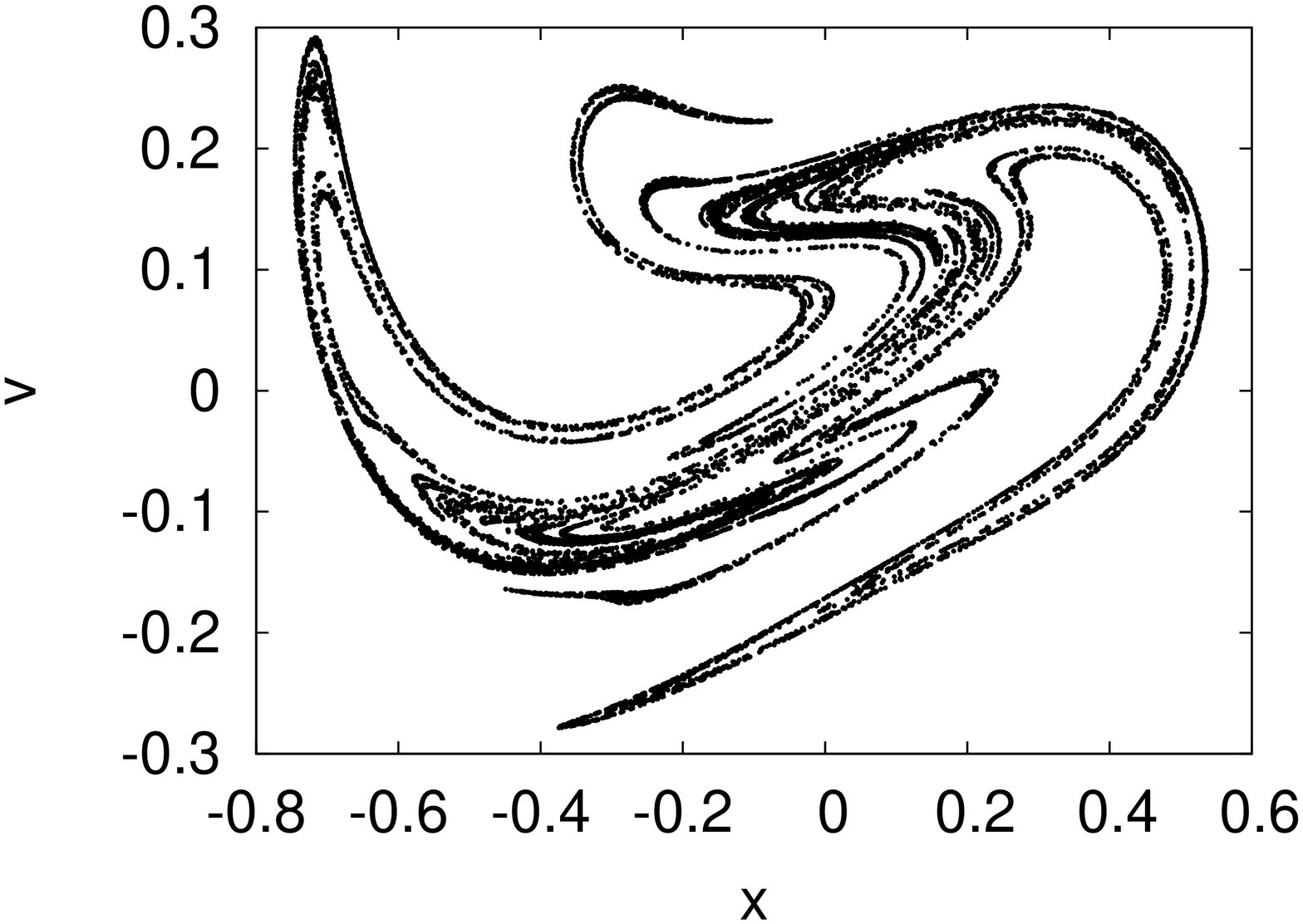,width=6cm,angle=-90}
\caption{\label{fig_attO2C_no} Explosion of the chaotic attractor for oscillator O2 with 
parameter set {\em (C)} and for increasing control-parameter values   $I_0=0.03476,~
0.03500,~0.03540,~0.03684$ 
 from the left to the right and from the top to the bottom, respectively.}
\end{figure*}
\end{center}

\begin{center} 
\begin{figure*}[htb]
\epsfig{file=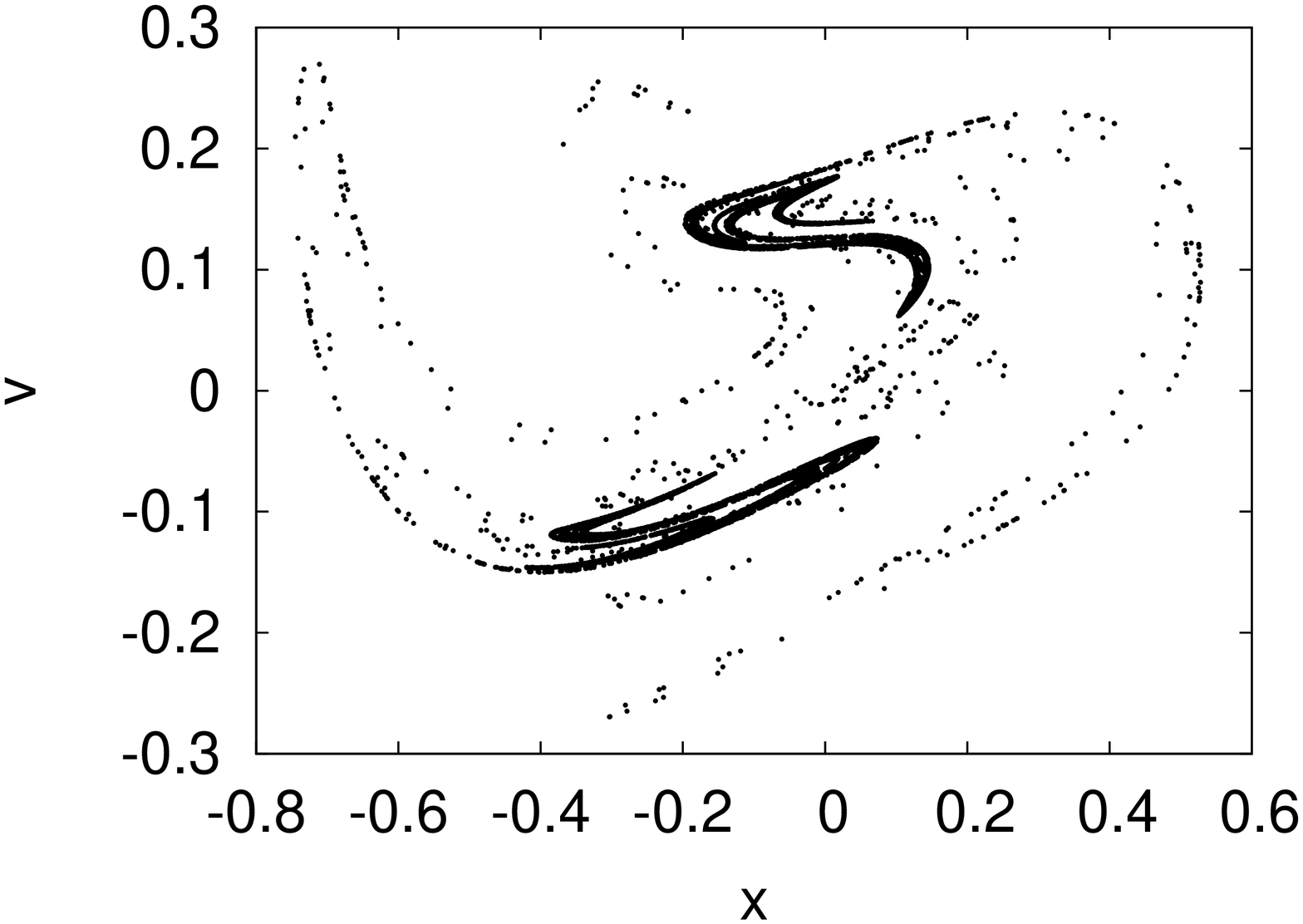,width=6cm,angle=-90}
\epsfig{file=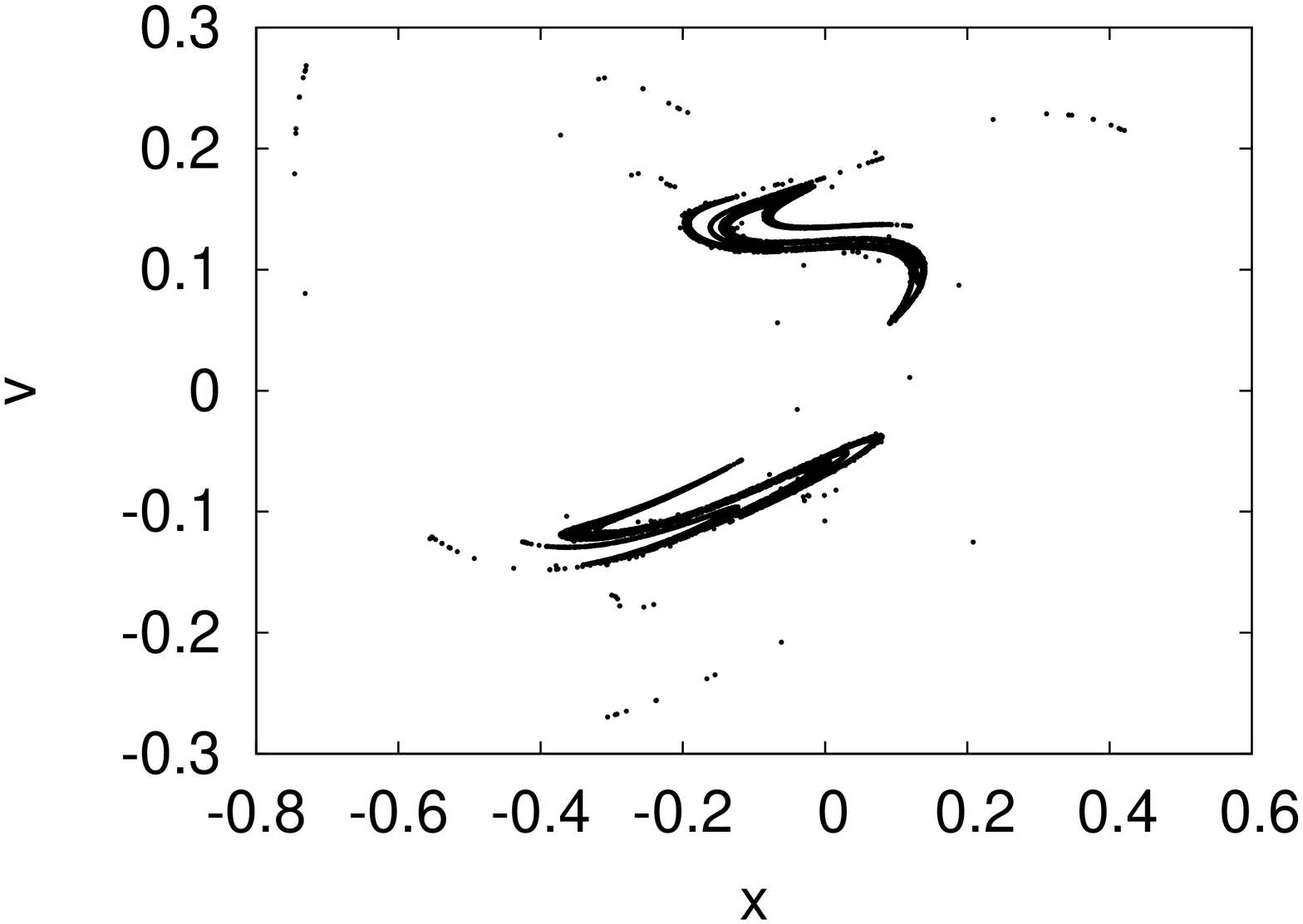,width=6cm,angle=-90}
\epsfig{file=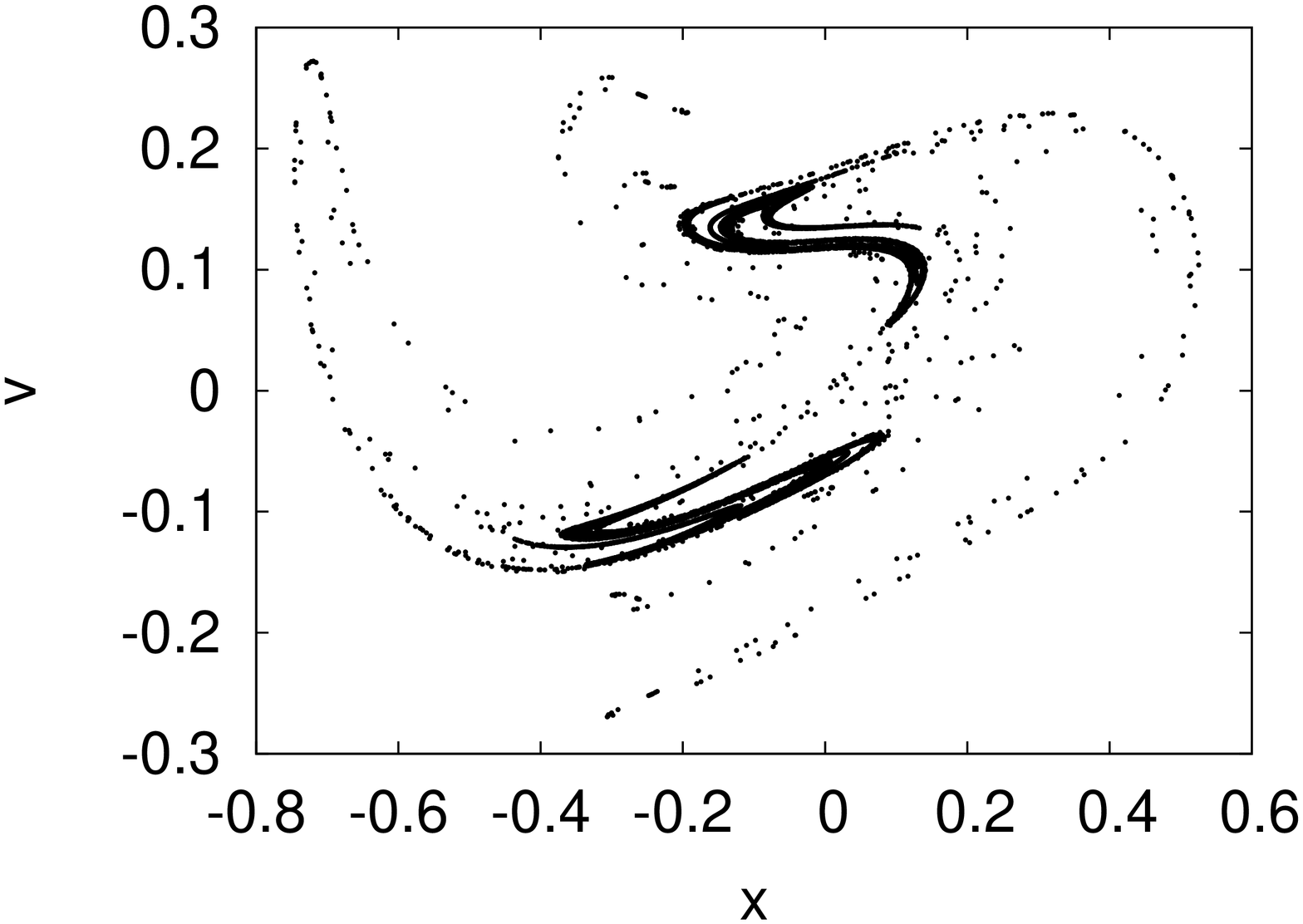,width=6cm,angle=-90}
\epsfig{file=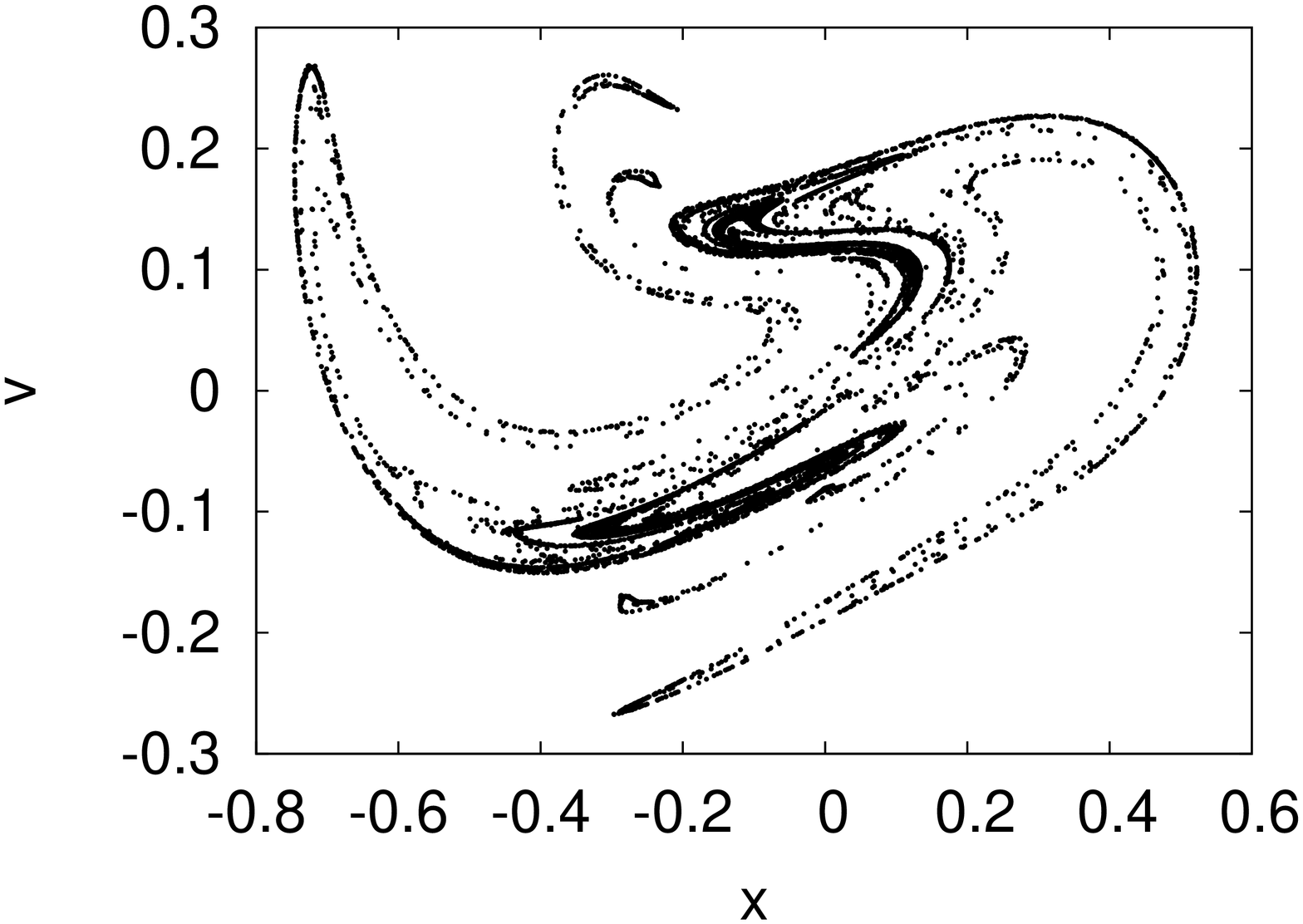,width=6cm,angle=-90}
\caption{\label{fig_attO2C_cs} Explosion of the chaotic attractor for oscillator O2 with 
parameter set {\em (C)} and for decreasing control-parameter values   $I_0=
0.03460,~0.03400,~0.03396,~0.03316$ 
 from the left to the right and from the top to the bottom, respectively.}
\end{figure*}
\end{center}

 Taking the set of pictures on the
strange attractor with these rather fine steps  of the control parameter, 
we have found that there is a critical value $I_{0c}\approx 0.03476$, where the attractor
can be considered  unexploded, consisting of two bunches separated by a phase-space
region of  very low population, but not being empty as compared to the corresponding
 phase-space regions belonging to the unexploded strange attractors found by us for
 oscillator O1.  This can be seen from the comparison of
  the top left picture in Fig. \ref{fig_attO2C_no}
to the top left pictures in Figs. \ref{fig_attO1A} and \ref{fig_attO1B}, when one keeps in
 mind that 
all pictures  of the Poincar\'e maps represent a statistical sampling of similar quality
of the distribution of the trajectory points on the strange attractor.
The serii of attractors in Figs.  \ref{fig_attO2C_no}
 and \ref{fig_attO2C_cs} show that both with increasing and decreasing values of the
control parameter $I_0$ the attractor starts to build up with more and more points
in the phase-space region separating the highly populated  bunches. 
With increasing values of the control parameter $I_0>I_{0c}$  the gradual building up of
 the attractor seems to continue until the value $I_0\approx 0.038$ is reached where
a wide regular window opens up (c.f. the negative average Lyapunov exponents in Fig.
\ref{fig_lambO2C}.) Two isolated points of regular behavior has also been 
detected 
in the interval $I_0\in \lbrack 0.03476,~0.038\rbrack $ but those seem not to influence 
the explosion process.
 In the direction of decreasing $I_0<I_{0c}$ the strange
 attractor starts to build up again, for $I_0=0.0346$ (the most left picture on the top
of Fig. \ref{fig_attO2C_cs}) there are more points populating the phase-space region between the bunches than for $I_0=0.03476$ (the most left picture on the top
of Fig. \ref{fig_attO2C_no}), but then the set of pictures in Fig. \ref{fig_attO2C_cs}
shows that  for $I_0=0.03396$
the inner phase-space region becomes again almost empty, the explosion is suddenly
 interrupted, its result is almost erased and then a restart of the explosion leads to
the  gradual building up of the attractor till the control parameter falls off
to the value $I_0\approx 0.03316$. The sudden break up and restart of the explosion
seems to appear as a result of the short regular window around $I_0\approx 0.034$ (see the points with negative average maximal Lyapunov exponents in Fig. \ref{fig_lambO2C}).
 The explosion process seems then to continue until the upper edge of the wide regular window 
$I_0\in \lbrack 0.030,~  0.033\rbrack$ is reached.
The evolution of the attractor found by us
is in agreement with the observations reported in \cite{Ueda80}
(see Figs. 3 (a) and 
(b) for $I_0=0.030$ and $0.045$, respectively, and  Fig. 6 for $I_0=0.035$ in \cite{Ueda80};
$B=I_0$ in the author's notation). The case of explosion for $I_0>I_{0c}\approx 0.03476$
with increasing control parameter $I_0$ has been  thoroughly investigated in   \cite{Ueda80}
and also its dynamical explanation has been given. Namely, the explosion in that case 
arises when the stability and instability regions of a hyperbolic 
fixed point tangentially touch one another.
The explosion of the 
attractor for control-parameter values $I_0<I_{0c}$ decreasing from the critical value,
and its sudden interruption and restart are not mentioned  in   \cite{Ueda80}.

 \section{Maximal Lyapunov exponent}\label{Lyap}

\subsection{On the numerical algorithm and its sensitivity}

We have taken the decision on the regular or chaotic behaviour of the
trajectories according to the sign of  their maximal Lyapunov exponents for the determination of which we have developed a $C^{++}$ code implementing
 Bennetin's method \cite{Osel68,Chiri71,Casa76,Bene76,Cont78,Shim79,Bene80a,Bene80b,Wolf85}.
 The time evolution of the minute vector
 $\vec{\ell}(t)=(\xi(t),
\eta(t))$ attached with its bottom to the reference trajectory at the time $t=t_0(>t_d)$
  is then determined performing a sequential
 renormalization of its length to its original size
after each time interval $\tau$. The tip of the vector  $\vec{\ell}(t)$ points to the
  side-trajectory started at times $ t_k=t_0+k\tau$ in the $(k+1)$-th sequence. 
 The reference trajectory has been determined 
 as described in the previous section, the side-trajectories have been determined by
 solving the linearized equations of motion using  the same fourth-order Runge-Kutta
 algorithm and the same time step $\Delta t=0.001$. The linearized
equations of motion are given as
\bea
 {\dot \xi}&=&\eta,\nn
{\dot \eta} &=&-\gamma\eta -\delta^2\lbrack 1+ 3x^2(t_k') \rbrack \xi
\eea
for oscillator O1 and
\bea
{\dot \xi}&=&\eta ,\nn
{\dot \eta}&=&-\gamma\eta -3x^2(t_k') \xi
\eea
for oscillator O2,
 where $t_k'= t_k+t' $ with the time variable $t'\in \lbrack 0, \tau)$ started from zero  at
 the beginning of each sequence and the dot stands here for differentiation with respect to 
$t'$.
At the end of any $(k+1)$-th sequence the vector $\vec{\ell}(t_k+ \tau-\epsilon)$ has been
rescaled to its original size $\ell_0=|\vec{\ell}(t_0)|$ and the vector
\bea
  \vec{\ell}(t_{k+1})= \frac{ \ell_0}{|\vec{\ell}(t_k+ \tau -\epsilon)|}
\vec{\ell}(t_k+ \tau-\epsilon)
\eea
(with $\epsilon \to 0^+$) obtained in this manner  is used as initial condition for solving
 the linearized equations in the next sequence. Having determined the side-trajectories 
for the first $n$ sequences, we obtained the quantities
\bea
 k_n(\ell_0,\tau)&=&\frac{1}{n\tau}\sum_{k=1}^n 
\ln \frac{ |\vec{\ell}(t_k+ \tau-\epsilon)|}{\ell_0}
\eea
 which for large number $n$ of the sequences converge to the maximal
 Lyapunov exponent \cite{Osel68,Chiri71,Casa76,Bene76,Cont78,Shim79,Bene80a,Bene80b,Wolf85},
\bea
  \lambda &=&\lim_{n\to \infty} k_n(\ell_0,\tau).
\eea

\begin{center}
\begin{figure}[htb]
\epsfig{file=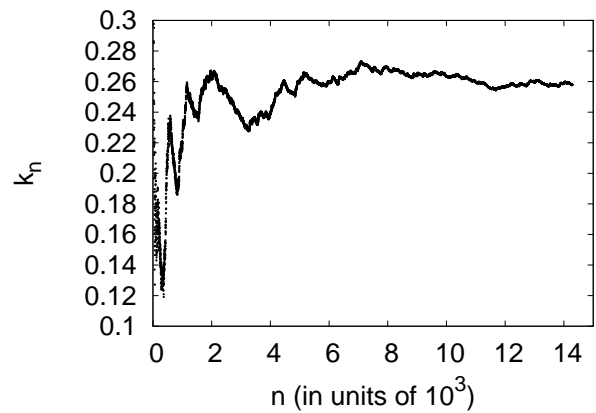,width=6cm,angle=-90}
\caption{\label{fig_lyapconv} Convergence of the $k_n$ values with increasing $n$ for
 oscillator O1 with  parameter set {\em (A)} for  $I=788.75$ and the choice $\tau=0.7$,
  $\xi(0)=\ell_0=10^{-9}$,  $\eta(0)=0$. }
\end{figure}
\end{center}

A detailed  analysis of the numerical algorithm has been performed in order to determine its
sensitivity to the number $n$ of the time steps $\tau$, the choice of the parameters
 $\ell_0$ and $\tau$, the orientation of the vector $\vec{\ell}(t_0)$, as well as to the
 initial conditions. The parameters $n\approx 15000$, $\ell_0=10^{-9}$, $\tau=0.5$,
 $\xi(0)=\ell_0$, $\eta(0)=0$ have been used for the determination of the maximal
 Lyapunov exponents. Fig.
\ref{fig_lyapconv} illustrates the typical convergence of the $k_n$ values. The
 maximal Lyapunov exponent has been determined as the mean of the last 20 $k_n$ values in
 the sequence which show up a few per cent variance. A similar sensitivity has been
 obtained for variing the duration $\tau$  of the time sequences taking the values 
$\tau=0.1,~0.2,~0.5,~0.7$, and for
variing the length $\ell_0$ of the initial minute vector within a factor of $\sim 10$. 
 We have also
rotated the initial vector $\vec{\ell}(t_0)$ by discrete angles $\Delta \varphi=\pi/4$
 turning around by the angle $2\pi$ in the $(\xi,\eta)$ plane, and observed that the 
last 20 $k_n$ values remained stable within a few per cent. Thus the algorithm with the
 choice of the parameters $n\approx 15000$, $\ell_0=10^{-9}$, $\tau=0.5$,
 $\xi(0)=\ell_0$, $\eta(0)=0$ has  provided fairly stable results. 

The maximal Lyapunov exponents determined in the above described manner turned out to be
 much more sensitive to the initial conditions $(x(t_0), v(t_0))$ determining the point on
 the reference
 trajectory, where the minute vector $\vec{\ell}(t_0)$ has been attached to it. Therefore,
we have determined maximal Lyapunov exponents averaged over several 
 initial conditions.
  As described in Sect. \ref{oscis} the reference trajectories have been followed up to 
$t\approx 10^5$. For
the various initial conditions such points of the corresponding Poincar\'e map, i.e., 
those of the attractor have been selected which belong to randomly chosen 5 to 10  integer
 $t_0$ values in the time interval  $\lbrack 5\cdot 10^4, 8\cdot 10^4\rbrack$. Generally
 more  initial conditions have been taken for values of the control parameter for which 
the chaotic attractor just explodes. It has been observed that the mean ${\bar \lambda}$ of
 the maximal
 Lyapunov exponents belonging to the various initial conditions exhibit a variance of about
 10 per cent for any  values of the control parameter involved in the investigation.

In order to test our numerical code for the determination of the maximal Lyapunov exponent
we applied it to the Van der Pol oscillator investigated in \cite{Ramasubramanian} where
the value $\lambda=0.0985$ is given in comparison to our result $\lambda=0.0992$
obtained for the same parameter values and initial condition. The discrepancy of 0.7 per cent is in agreement with the sensitivity of the algorithm to the various parameters.

\subsection{Maximal Lyapunov exponents of the exploding chaotic attractors}

\begin{center}
\begin{figure}[htb]
\epsfig{file=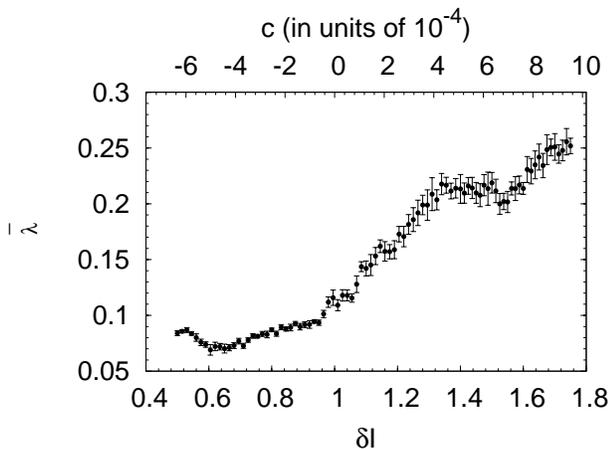,width=6.0cm,angle=-90}
\caption{\label{fig_lambO1A} The average maximal Lyapunov exponent ${\bar \lambda}$ vs.
 the shifted control parameter
 $\delta I=I-787.0$ for oscillator O1 with parameter set {\em (A)}. Also the scale $c=(I-I_c)/I_c$ with $I_c\approx 788.0$
is shown. }
\end{figure}
\end{center}

\begin{center}
\begin{figure}[htb]
\epsfig{file=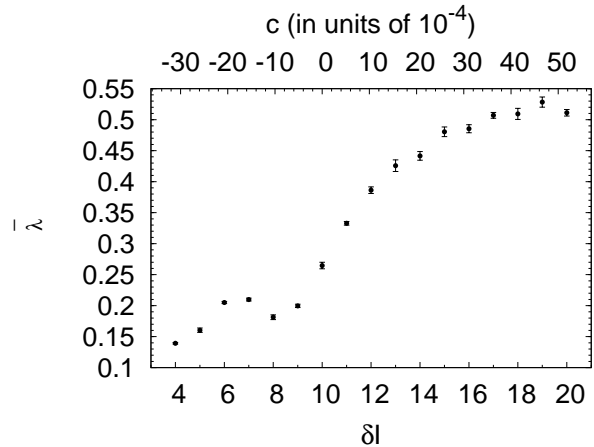,width=6.0cm,angle=-90}
\caption{\label{fig_lambO1B} The average maximal Lyapunov exponents ${\bar \lambda}$ vs. 
the shifted control parameter
 $\delta I=I-1920$ for oscillator O1 with parameter set {\em (B)}.  Also the scale $c=(I-I_c)/I_c$ with $I_c=1930$
is shown. }
\end{figure}
\end{center}

The dependence of the average maximal Lyapunov exponent on the control parameter
of the exploding strange attractor has been determined for oscillator O1 with parameter 
sets $(A)$ and $(B)$ and for oscillator O2 with parameter set $(C)$.
An approximately linear raise of the average maximal Lyapunov exponent $\bar{\lambda}$
during the explosion of the strange attractor with increasing values of the control 
parameter $I$ has been observed
 for  oscillator O1 for both parameter sets {\em (A)} and {\em (B)}, as shown in Figs.
 \ref{fig_lambO1A} and \ref{fig_lambO1B}, respectively. Here the points represent 
maximal Lyapunov exponents averaged over 10 initial conditions in the region of the
 explosion. In these cases the explosion is not interrupted by windows of regular behaviour.

A comparison of the graph in Fig. \ref{fig_lambO1A} with
the plots in Fig. \ref{fig_attO1A} 
reveals the following behaviour. For $I=787.91$ (and similarly in the interval $I\in \lbrack 787.4, \sim 787.9\rbrack$) the chaotic
 attractor 
consists of three  disjoint bunches sparated by an empty phase-space region, while
the average maximal Lyapunov exponent remains almost constant.
When the burst out of the attractor into the phase-space region between the highly populated
 bunches sets on at $I=I_c\approx 788.0$,   an approximately linear raise
of the average maximal Lyapunov exponent starts with increasing values of the control
 parameter $I$  and lasts in the whole interval  $I\in \lbrack 788.0, 788.3 \rbrack$, where
 the explosion takes place. Finally, the $\bar \lambda$ values saturate
 for $I \in \lbrack 788.4,\sim 788.5\rbrack$, where the explosion stops. 
The average maximal Lyapunov exponent raised by a factor of $\sim 3$ during the explosion.

A similar approximately linear  increase of the average  maximal Lyapunov exponent
$\bar \lambda$  with increasing control
 parameter $I$ can be recognized in Fig. \ref{fig_lambO1B}. The comparison of 
 \ref{fig_lambO1B} with the series of plots in Fig. \ref{fig_attO1B} reveals
 again that the linear raise of ${\bar{\lambda}}$ accompanies the explosion of the strange 
attractor.
 For $I\in \lbrack 1926, \sim 1929\rbrack$, when 
 the yet unexploded strange attractor
 consists of three disjoint bunches,  the average maximal Lyapunov exponent does not change
significantly.
In the interval  $I\in \lbrack \sim 1930, \sim 1935\rbrack$ where the strange attractor
 explodes 
there occurs an almost linear raise of $\bar \lambda$ again.
 Finally, the  ${\bar \lambda}$ values saturate for $I\in \lbrack \sim 1935, 1940\rbrack$
where the explosion of the attractor stops.
 The average maximal Lyapunov
 exponent raised  by a factor of $\sim $2.5 in this case.

\begin{figure}[ht]
\epsfig{file=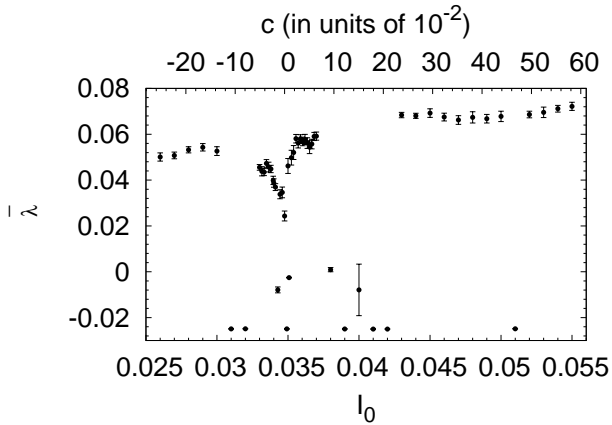,width=6.0cm,angle=-90}
\caption{\label{fig_lambO2C} The average maximal Lyapunov exponent ${\bar \lambda}$ vs. 
the  control parameter $I_0$ for oscillator O2 with parameter set {\em (C)}. Also the scale
$c= (I_0-I_{0c})/I_{0c}$ with $I_{0c}=0.03476$ is shown. }
\end{figure}

For the oscillator O2 with the parameter set {\em{(C)}} we 
recovered the same windows of the regular behaviour breaking time-to-time the explosion of
 the chaotic attractor
with increasing control parameter $I_0$  observed in \cite{Ueda80}
 previously.
This was realized by making a point-to-point comparison with Fig. 7 in \cite{Ueda80}
 showing the exponent-like quantities determined there.
Even the few points where the author could not uniquely decide whether the motion is 
regular or chaotic reappeared in our calculation showing up both positive and negative
 $\lambda$ values for the various initial conditions. These points generally are such
 where the system is 
rather close to an edge-of-chaos system.
 Our results on  the maximal Lyapunov exponents ${\bar \lambda}$ averaged over the
 various initial conditions are shown in Fig. \ref{fig_lambO2C}.
Here the  trajectory started at $x(0)=v(0)=0.1$ has been used to obtain the chaotic 
attractor on which additional  initial conditions for 10 reference trajectories have
 been chosen, in order to evaluate the average exponent.

Here we identified the  critical value with $I_{0c}\approx 0.03476$ at which
we have found the smallest positive value of the average maximal Lyapunov exponent
 ${\bar \lambda}_c\approx 0.0243$.
In between the wider regular windows $I_0\in \lbrack  0.030,~ 0.033 \rbrack$
on the left and $I_0\in \lbrack  0.038,~  0.043\rbrack$ on the right the average
 exponent ${\bar \lambda}$ shows up a rapid increase from its minimal value
when $I_0$ moves away from its critical value in both directions and a kind of saturation
occurs close to the edges of the wide regular windows. Such a matter-of-fact
reflects the explosion of the strange attractor for $I_0$ values tuned away from its critical value as shown in Figs.   \ref{fig_attO2C_no} and \ref{fig_attO2C_cs} and discussed in Sect. 
\ref{oscis}.
 The overall trend of raising
is approximately linear in both directions, again accompanied by the total increase of
${\bar \lambda}$
with the factor of $\sim 2.0$
and $\sim  2.5$  for decreasing and increasing $I_0$ values, respectively.
 Under the present accuracy of our 
calculations such an overall trend seems not to be disturbed 
 by the rather short regular windows breaking the chaotic behaviour in
the interval $I_0 \in \lbrack  0.33,~  0.38\rbrack$. 
Neither the structural change of the strange attractor (mentioned in connection with
the plots in Fig. \ref{fig_attO2C_cs}) breaks the trend of  variation
of the average maximal Lyapunov exponent when the control parameter
takes the values
 $I_0=0.03460,~0.03400,$ and
$0.03396$ and swaps a rather short regular window in the close neighborhood of
$I_0=0.03428$.

\section{Generalized dimensions}\label{gendim}

\subsection{Determination of the generalized dimensions}

The generalized dimensions $D_q$ characterize the multifractal structure of the strange
 attractor \cite{Schuster05,Hilborn,Beck95}. We have chosen the sandbox method which provides the generalized dimensions
via the slope of the logarithm of the so-called generalized correlation sums $C_q$ vs. the
 logarithm of the box size \cite{Schuster05,Hilborn}. Let $P_i$ $(i=1,2,\ldots, N)$ be a series of
 points on the multifractal under consideration given via the vectors ${\vec r}_i$ in
 a $d$-dimensional Euclidean space.
 Hyperspheres of radii $R$ are centered at each of the points  $P_i$
$(i=1,2,\ldots, N)$ and the relative frequencies 
\bea
w_i(R)&=& \frac{1}{N-1}\sum_{j=1, j\not=i}^N
  \Theta (R-|\vec{r}_i-\vec{r}_j|)
\eea
  of finding in those another  point $P_j~(j\not =i)$ of the series  are determined.
For $q\not= 1$ the average of the
powers $w_i^{q-1}(R)$ are  taken over the series of points in order to determine  the
 generalized correlation sums 
\bea
  C_q(R)& = &  \frac{1}{N} \sum_{i=1}^N w_i^{q-1}(R).
\eea
Let us remind that for $q>1$ integers and $N\gg1$ the sum $ C_q(R)$ represents the average
 probability to find a number of $q$ points in a hypersphere of radius $R$.
 The generalized dimensions are  read off from the
 asymptotic scaling $C_q(R)\sim R^{(q-1)D_q} $ for $R\to0$,
\bea
  D_q&\sim & \frac{1}{q-1} \frac{\ln C_q(R)}{\ln (R/R_{max})},~~q\not=1
\eea
where $R_{max}$ is the maximal distance of the points in the  series. 
The dimension $D_1$, the information dimension is obtained as the limit $D_1=\lim_{\epsilon\to 0} D_{1+\epsilon}$
that yields
\bea
  D_1&\sim & \frac{  (1/N)\sum_{i=1}^N \ln w_i(R)}{\ln (R/R_{max})} .
\eea
It is well-known that $D_2$ and $D_0$ are  the correlation and the Hausdorff dimensions,
 respectively. The latter should be smaller than the dimension $d_{ph.sp.}$ of the phase-space ($D_0<d_{ph.sp.}=3$ in our case) because of the not space-filling fractal structure of the strange attractor. Furthermore, the generalized dimensions are monotonically decreasing with increasing
value of the parameter $q$, i.e., it holds the inequality $D_q\ge D_{q'}$ for $q<q'$.  The sum $C_q(R)$ is
 dominated by the regions of the strange attractor with large and small
occupation probabilities, respectively, for   $q\ge 0$ and
$q<0$. Therefore one expects that the generalized dimensions $D_q$ for negative  parameter values
 $q<0$ get large when the strange attractor exhibits extended regions of low occupation 
probability. This is the feature we shall use to characterize the explosion of the strange
 attractor.

We have used  Grassberger and Procaccia's correlation
 sum approach \cite{GrassPro83b,GrassPro83,Grass83,BenMi84,Farme83,Hilborn} when
 the time series of a single variable measured on the attractor is embedded first
into a $d$-dimensional vector space and then the correlation sum is evaluated from 
the series of vectors in the embedding space.
For a proper choice of the dimension $d$ of the embedding space
 the embedded trajectories will have the same geometric and dynamical properties
as the true trajectory has in the  phase space. We constructed a number $N$ of
 $d$-dimensional vectors of the
 embedding space  from the time series of the coordinate variable $x_n=x(t_n)$ with
 $n=1,2,\ldots, N_a$ `measured' on the Poincar\'e map of the attractor. The series of vectors
\begin{eqnarray}
  {\vec r}_j&=& (x_j, x_{j+1}, \ldots, x_{j+d-1})
\end{eqnarray}
$(j=0,1,\ldots, N-1)$ have been constructed where  we have  chosen $t_0\approx 10^5$ 
and the time interval between the sampled values as well as
the lag time between the successive vectors has been set equal to the time  period 
 of the periodic driving force.
The trajectories started from the point $(x(t_0),v(t_0))$
of the strange attractor  have been determined numerically as described in Sect. \ref{oscis}. 
The embedding space has been endowed by  the Euclidean distance.
The various parameters of the embedding algorithm were settled on the strange attractor for
oscillator $O1$ with parameter set $(A)$ for the  value $I=788.22$ of the control parameter,
 i.e.,
for a strange  attractor  in the `midway' of being exploded. 

The determination of the asymptotic scaling region of the correlation sums is the  basic
 ingredient of Grassberger and Procaccia's method. The double-logarithmic plots $\ln C_q $ vs. $\ln R$ have been taken by the steps $\Delta \ln R \approx 0.3$.  
 Asymptotic scaling generally occurs in the
 interval $R_l\le R\le R_u$ where $R_u\stackrel{<}{\sim} R_{max}$
and $R_l> R_{min}$ with the estimated lower bound 
$R_{min}$. When data resulted from numerical
computations are used - like in our case - $R_{min}$ can be estimated in
 terms of the bit resolution $b$ by which the data
are represented in the computer as $R_{min} \sim 2^{-(b-2)}R_{max} \sim 10^{-4}R_{max} $ for
$b=15$, i.e., for double precision computations  
  \cite{Hilborn}.  According to our numerical experience the lengths of the scaling intervals
are different for the various strange attractors and the various choices of the parameter $q$
and had to be determined in each particular case separately.
 The relatively small number of 
embedded vectors  used (c.f. the discussion below) resulted in a statistical noise causing 
 almost constant tails in the double-logarithmic plots $\ln C_q $ vs. $\ln R$ and the actual
 value of $R_l$ exceeded generally  the estimated lower bound $R_{min}$ with its rather typical values $R_l/R_{max}\sim 10^{-3} - 10^{-2} $.
 Although we were able to identify the  scaling  regions in each of the cases,
 there occurred  some ambiguity as to their boundaries $R_l$ and $R_u$.
 The statistical errors of the generalized dimensions
 $D_q$ presented by us below include the error of the fit of a
 straight line to the log-log plot in the scaling region
 and the generally even larger error from the somewhat
 ambiguous choice of the scaling region. The latter error has been estimated through
the variance of $D_q$ values obtained for various, slightly different choices of the
scaling region.

 Recording a number $N=5\cdot 10^4$ of vectors we investigated the dependence of the various
generalized dimensions $D_q$ for $q=-4,-3,\ldots, 3,4$ on the dimension $d$ of the 
embedding space in the range $d\in \lbrack 3 , 6\rbrack$. It was established that 
 the $D_q$ values for $q>0$
  are relatively stable although slightly raising with the increase of the embedding
 dimension  (see Table \ref{tab_Dq_d}), while
for $q<0$ they blow up with increasing $d$.  For our analysis
we have chosen $d=5$ for the dimension of the embedding space. This  is in accordance
 with
the observation \cite{Hilborn} that for a dissipative system
the choice of $d$ at about twice the fractal dimension of the attractor can be sufficient
to mimic the dynamics on the attractor. For positive parameter values $q>0$ one would expect the 
saturation of the $D_q$ values with the further increase of the dimension $d$, but then in our 
case the  relatively small number  $N$ of the vectors would result in a very low probability to
 find another vector in the neighbourhood of radius $R_{min}$ of any given vector  \cite{Hilborn}. 
In order to get at least $2$ vectors in average in a neighborhood of radius $R_{min}$ we would
 need the number $N\sim 2\cdot 2^{(b-2) d}\sim 10^{20}$ of vectors for $b=15$ and $d=5$. 
For $N\sim 5\cdot 10^4$,  $b=15$ and $d=5$ we get  $N  ( R_{max}/R_l )^{-d}\sim 1/2$ points in a
 neighborhood of radius $ 10^{-1} R_{max} $. Therefore the statistics we have is rather poor, but
a significant increase of the number $N$ of the vectors with several orders of magnitude is also
 not available. The poor statistics explains why the scaling intervals $R_l\le R\le R_u$ were
found by us in some cases rather short and their endpoints somewhat  ambiguous.

\begin{table}
\begin{center}
\begin{tabular}{|c|c|c|c|c|c|}
\hline
 \multicolumn{1}{|c|}{$q$} & \multicolumn{4}{c|}{$D_q\pm \Delta D_q$}\cr
\cline{2-5}
   & $d=3$ & $d=4$ & $d=5$ & $d=6$ \cr
\hline
 $-4$ & $3.5\pm 0.5 $ & $2.2\pm 0.5$ &  $2.8\pm 0.5$ & $3.8\pm 0.5 $  \cr
\hline
 $-2$ &  $2.8\pm 0.4$ & $2.1\pm 0.4$ & $2.7\pm 0.4$ & $3.8\pm 0.4$  \cr
\hline
 $0$ & $1.9\pm 0.3$ & $2.1\pm 0.3$ & $2.2\pm 0.2$ & $2.3\pm 0.3$\cr
\hline
 $2$ & $1.1\pm 0.1$ & $1.1\pm 0.1$ & $1.05\pm 0.08$ & $1.2\pm 0.1$  \cr
\hline
 $4$ & $1.03\pm 0.04$ & $1.05\pm 0.04$ & $0.99\pm 0.03$ & $1.08\pm 0.04$  \cr
\hline 
\end{tabular}
\caption{\label{tab_Dq_d} Dependence of the generalized dimensions $D_q$ on the embedding dimension $d$, the errors include 
those of the fit of a straight line to the curve $\ln C_q$ vs. $\ln (R/R_{max})$ and
the ambiguity of the scaling interval.} 
\end{center}  
\end{table}

Also the dependence of the algorithm on the number $N$ of the vectors involved in the computation
\begin{table}
\begin{center}
\begin{tabular}{|c|c|c|}
\hline
\multicolumn{1}{|c|}{$q$} & \multicolumn{2}{c|}{$D_q\pm \Delta D_q$}\cr
\cline{2-3}
   & $N=2.5\cdot 10^4$ & $N=5\cdot 10^4$\cr
\hline
 $-4$ & $3.2\pm 0.5$ & $2.8\pm 0.5$   \cr
\hline
 $-2$ &  $3.0\pm 0.5$ & $2.7\pm 0.4$   \cr
\hline
 $0$ & $2.2\pm 0.3$ & $2.2\pm 0.2$ \cr
\hline
 $2$ & $1.06\pm 0.11$ & $1.05\pm 0.08$  \cr
\hline
 $4$ & $0.91\pm 0.04$ & $0.99\pm 0.03$   \cr
\hline 
\end{tabular}\caption{\label{tab_Dq_Nv} Dependence of the generalized dimensions $D_q$ on the number $N$ of the vectors
in the embedding space involved in the evaluation of the correlation sums. The errors are the same as in Table \ref{tab_Dq_d}.}
\end{center}  
\end{table}
has been investigated. It has been established that the correlation dimension $D_2$
remains stable within the estimated errors for $N=2.5\cdot 10^4, ~5\cdot 10^4,~7.5\cdot 10^4,$ and  $10^5$. The generalized dimensions $D_q$ for $-4\le q\le 4$ have been computed
 for $N=2.5\cdot 10^4$ and $5\cdot 10^4$. Table \ref{tab_Dq_Nv} shows that for $q\ge 0$
the $D_q$ values are identical within the errors in both cases, while for $q<0$ the increase of the factor of 2 of the number of vectors leads to a slight fall of the corresponding $D_q$ values but still within the errors. 
 Because the computer time for the evaluation of the correlation sums
 runs with $N^2$, we have chosen $N=2.5\cdot 10^4$ for the systematic investigations.

The numerically evaluated values of the generalized dimensions are rather sensitive to the
phase  of the periodic driving force at which the Poincar\'e sections are taken. We have performed
calculations for phase shifts $\Delta \phi=0,~\pi/2,~\pi, ~3\pi/2$. Typical values for the 
oscillator {\em O1} with parameter set {\em (A)} for $I=788.22$, $d=5$ and $N=2.5\cdot 10^4$ are shown in Table
\ref{tab_Dq_phase}.
\begin{table}
\begin{center}
\begin{tabular}{|c|c|c|c|c|c|}
\hline
\multicolumn{1}{|c|}{} & \multicolumn{4}{c|}{} &\multicolumn{1}{c|}{} \cr
\multicolumn{1}{|c|}{$q$} & \multicolumn{4}{c|}{$D_q\pm \Delta D_q$} &\multicolumn{1}{c|}{
${\bar D}_q\pm \Delta {\bar D}_q$} \cr
\cline{2-5}
   & $\Delta \phi=0$ & $\Delta \phi=\pi/2$ & $\Delta \phi=\pi$ & $\Delta \phi=3\pi/2$ &\cr
\hline
 $-4$ & $3.2\pm .5$ &  $ 5.5\pm .7$ & $3.3\pm .5$ & $5.4\pm .7$     & $4.0\pm .3$\cr
\hline
 $-2$ &  $3.0\pm .5$ & $ 4.7\pm .6$ & $2.9\pm .5$   & $4.6\pm .6$ & $3.6\pm .3$ \cr
\hline
 $0$ & $2.2\pm .3$ & $2.4\pm .3$ &$2.1\pm .3$  & $ 2.4\pm .3$ & $2.3\pm .1$\cr
\hline
 $2$ & $1.06\pm .11$ & $ 1.02\pm .11$ &$1.04\pm .11$ & $1.05\pm .11$ & $1.04\pm .05$\cr
\hline
 $4$ & $0.91\pm .04$ & $0.87\pm .04$ &$ 0.98\pm .04$ &$0.86\pm .04$ &$ 0.91\pm .02$\cr
\hline 
\end{tabular}\caption{\label{tab_Dq_phase} Dependence of the generalized dimensions $D_q$ on the 
phase shift $\Delta \phi$ by which the Poincar\'e sections are taken. 
 The errors $\Delta D_q$ are the same as in Table \ref{tab_Dq_d}, ${\bar D}_q$ and 
$\Delta {\bar D}_q$ are the weighted averages
over the phase shifts.
}
\end{center}  
\end{table}
As a rule, the changes of the generalized dimensions $D_q$ with the phase shift $\Delta \phi$ 
exceed the estimated errors $\Delta D_q$.  As to the investigated exploding
attractors of oscillator $O1$, it has been observed that a phase shift of
 $\Delta \phi=\pi$ corresponds to an almost rigid rotation of the
Poincar\'e map by the angle $\pi$, although the phase shifts with intermediate values
 $0<\Delta \phi<\pi$ involve the distortion of the strange attractor, as well. This explains
that the generalized dimensions of the system determined from Poincar\'e sections taken
with a phase difference $\pi$ are identical within their errors, but those belonging to
Poincar\'e sections taken with the phase shift $\pi/2$ deviate much more than their statistical
errors. For the discussed exploding attractor of oscillator $O2$,
the phase shift seems always to cause a combination of some rigid rotation and distortion
 of the strange attractor.
 Due to this state of affairs we have determined the weighted averages ${\bar D}_q$ of the
 generalized dimensions belonging to the Poincar\'e sections taken with phase shifts $0$,
 $\pi/2$,
$\pi$, and $3\pi/2$ and these average values have been used for the characterization of the 
strange attractors. 

One may be cautious about the ability of our procedure to yield
 the exact values of  $D_q$'s basicly due to the poor statistical sampling of the attractor,
i.e., due to the rather small number $N$ of the  vectors in the embedding space. Nevertheless,
the obtained  values of the generalized dimensions enable one for making comparisons
of the strange attractors of a given oscillator when the control parameter gradually
changes. The $q$-dependence of the generalized dimensions
for any of the investigated particular systems follows the theoretically expected
monotonically falling off tendency with increasing parameter $q$. As a rule
 the $D_0$ values satisfy the inequality $2<D_0<3$ which means that the strange
 attractor does not fill the 3-dimensional phase space. Although the $D_q$ values for
 $q<0$ may be far away of their exact values, but for all attractors they are determined
 by the same algorithm and in that manner their changes should be characteristic for 
the alteration of the fractal structure of the exploding strange attractor.

\subsection{Explosion of the strange attractor and the generalized dimensions}

\begin{figure}[ht]
\epsfig{file=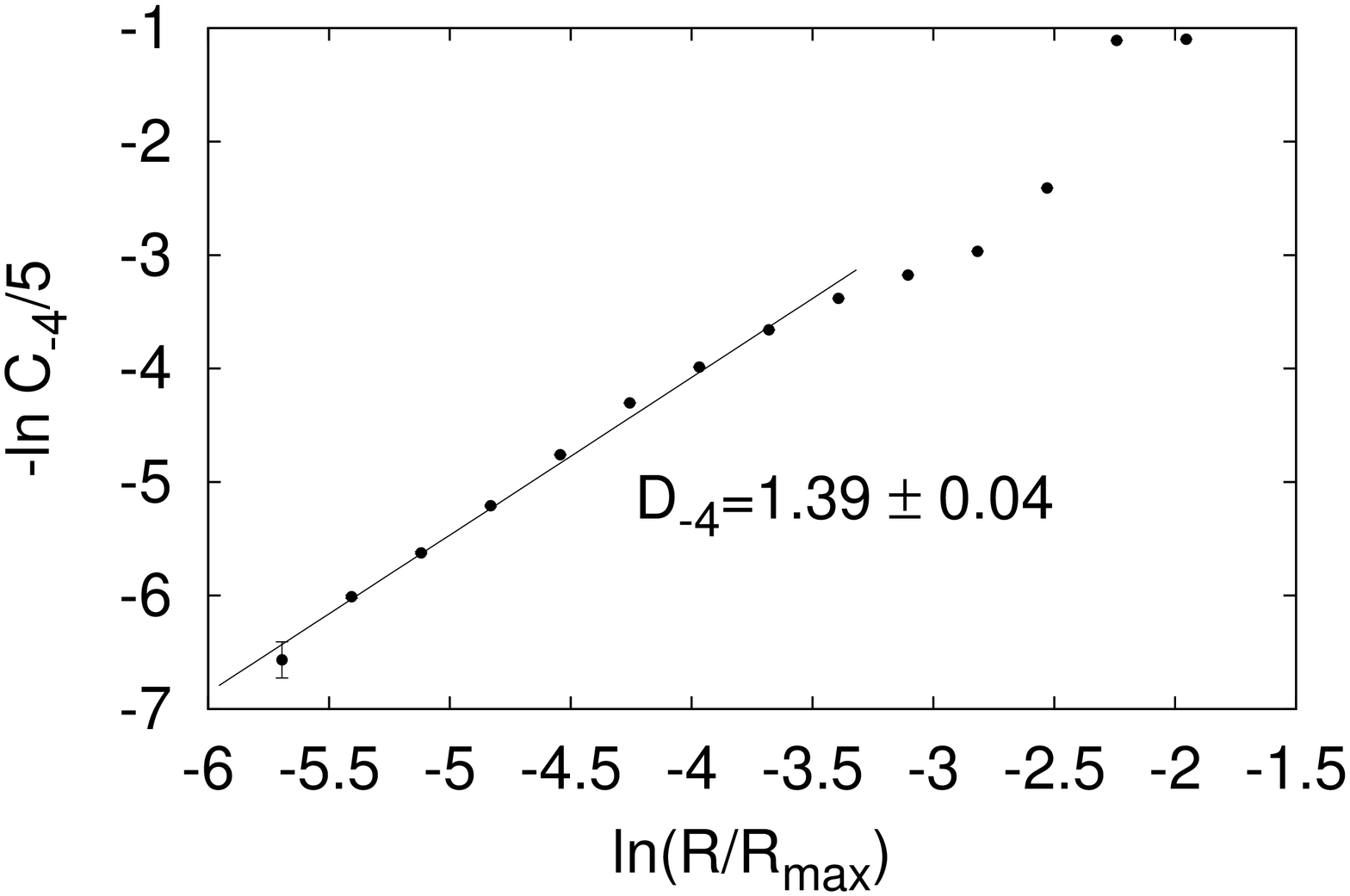,width=6.0cm,angle=-90}
\epsfig{file=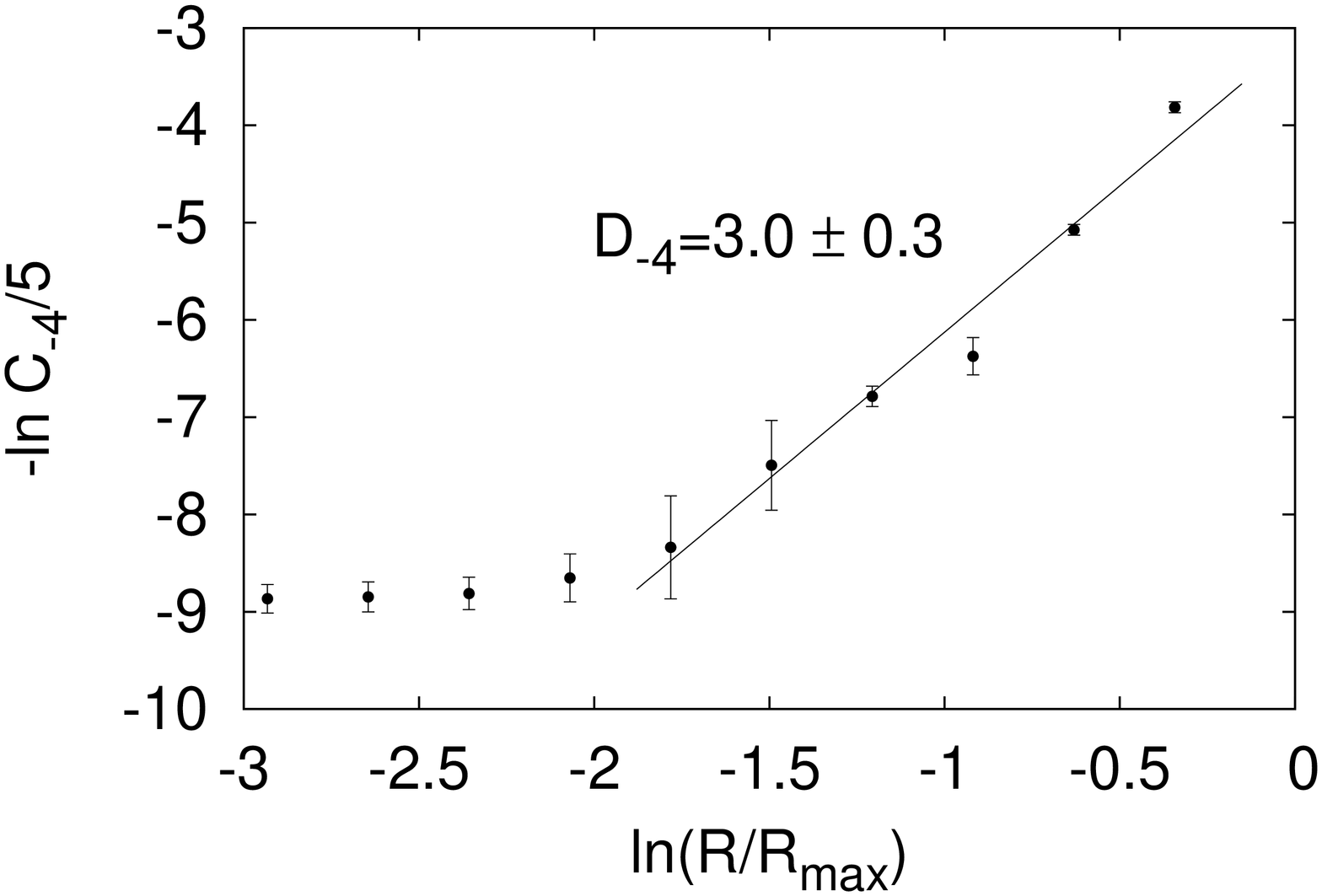,width=6.0cm,angle=-90}
\caption{\label{fig_lnCm4} Logarithm of the correlations sum $-\ln C_{-4}/5$ vs. $\ln (R/R_{max})$
for oscillator O1 with parameter set {\em (A)} for the strange attractor just before the
explosion for $I= 787.92$ (top) and just after it has been started for $I=788.01$ (bottom) with phase shift $\Delta \phi=0$. 
 }
\end{figure}

Before going into the detailed discussion of the results let us emphasize two features
of the generalized dimensions which are of particular importance in characterizing the
explosion of the chaotic attractor. It is well-known that the correlation sums for positive 
and negative parameters $q$ are dominated by the contributions of the densely and rarely
occupied regions of the attractor, respectively. Therefore, one expects that dimensions
$D_q$ with negative $q$ values for which $D_q$ approaches
already  the limiting value $D_{-\infty}$ are rather sensitive to the explosion of the attractor.
 When the strange 
attractor has not yet been exploded and consists of several bunches of points separated by an
 empty phase-space region one should get  much smaller values for the dimensions  $D_{-|q|}$ 
with $|q|\gg 1$ than the  values obtained after the sampled points of the trajectory
start  to occupy the region separating the highly populated bunches of the strange attractor.
 Therefore one expects  a
 sudden jump in the values of $D_{-|q|}$ for $|q|\gg 1$. Moreover, the 
$x$ coordinates of the points on the Poincar\'e map of the  unexploded attractor
represent a data set with gaps, i.e., these data belong to disjoint intervals. Consequently,
the correlation sums are expected to show up two different scaling regions one for
small and one for large separations of the point-pairs \cite{Hilborn}. For small 
$R_l\le R < R_b$ the pairs of points contributing to the sum $C_{-4}(R)$ belong to the same bunch
 of the attractor, while for large $R_s< R<R_u$  each point of the pairs belong to different bunches. Here $R_b$ and $R_s$ are, respectively,  the characteristic size of a single bunch and
that of the separation distance of the various bunches.
 After the explosion
 has been started, the gap in the data disappears and one obtains a single scaling region.
Typical scalings of $\ln C_{-4}$ vs. $\ln (R/R_{max})$ are shown in Fig. \ref{fig_lnCm4} for
oscillator O1 with parameter set {\em(A)}. Just before the explosion of the attractor there are 
two scaling regions: a long one for small point separations  and a rather short one for large 
separations, but  only a single scaling region appears when  the explosion process has already
 been set on work. When the explosion process proceeds the originally empty phase-space region
 between the bunches becomes more and more occupied by trajectory points and, consequently,
the two scaling regions  merge into a single one.
 Similar behaviour has been observed for oscillator O1 with parameter set
 {\em (B)}.
For oscillator O2 with parameter set {\em (C)} no double scaling has been observed for 
 values  $I_0\in \lbrack 0.033, 0.038\rbrack $, i.e., in between the left and right wide regular windows.   This is a consequence of not having an empty
phase space region between the highly populated bunches of the strange attractor,
 i.e., that of not having a gap in the data
neither for $I_{0c}=0.03476$ (where the  maximal Lyapunov exponent take its smallest value)
nor for $I_0= 0.03400$ (where a sudden break and restart of the explosion has been observed).
For cases when two scaling regions occurred for oscillator O1, we have determined the $D_{-4}$
value from the much longer scaling region for small separations. Thus,  for the exploding 
attractor the $D_{-4}$ values have been enhanced via the contributions of the separations $R$ 
of the order  of the size of the underpopulated region, as compared to
 the unexploded attractor for which  the contributions of
 separations not exceeding the size of the bunches dominate $\ln C_{-4}$ and 
yield a suppressed $D_{-4}$ value.
We have to mention that the second scaling regions for large separations are generally rather
short and not available for a reliable determination of another scaling dimension due to the low
statistics (due to the relatively small number of embedded vectors) in our computations. 
Nevertheless, its scaling exponent can be estimated generally a few times larger than the
 dimension $D_{-4}$ determined from the much longer scaling regions for small separations.

\begin{figure}[ht]
\epsfig{file=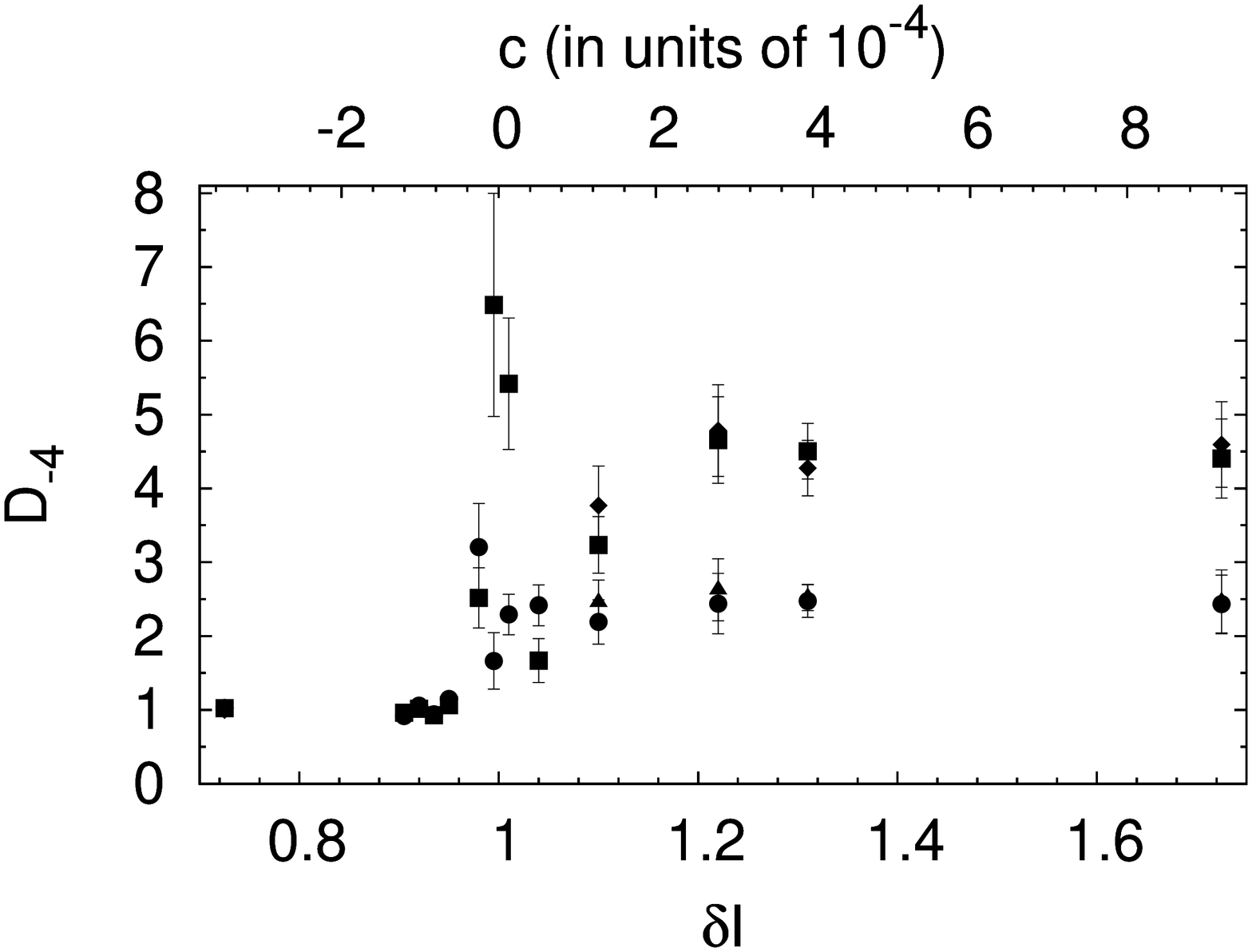,width=6.0cm,angle=-90}
\epsfig{file=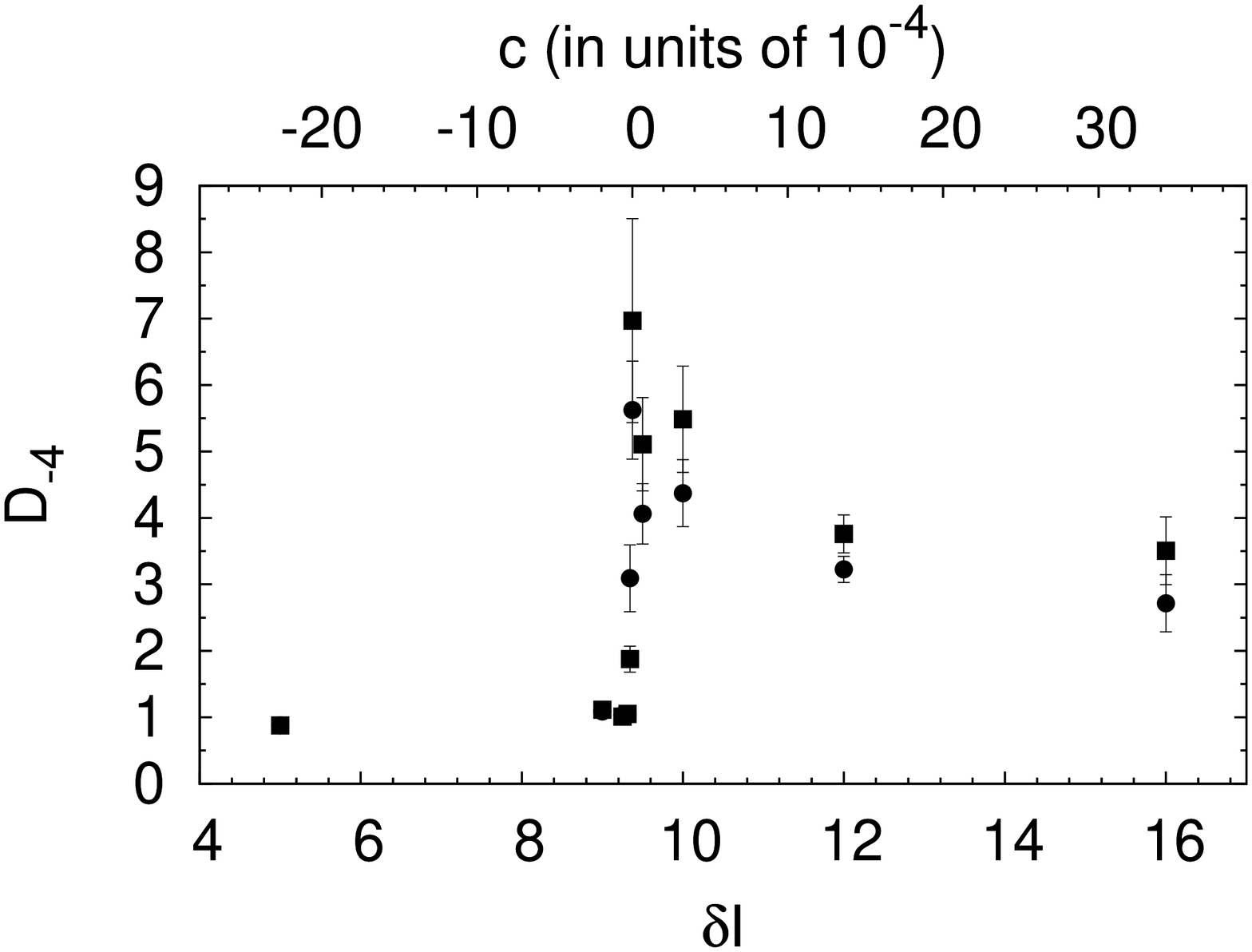,width=6.0cm,angle=-90}
\caption{\label{fig_Dm4O1phi} The generalized dimension $D_{-4}$ vs. the shifted control
parameter $\delta I$ (as given in Figs. \ref{fig_lambO1A} and \ref{fig_lambO1B}, respectively) for the strange attractor belonging to oscillator O1 with parameter sets
{\em (A)} (top) and {\em (B)} (bottom) for various phase shifts: $\Delta \phi=0$ (circle),
$\pi/2$ (square), $\pi$ (triangle), and $3\pi/2$ (diamond). Also the scale $c=(I-I_c)/I_c$ is shown.
 The $D_{-4}$ values are normalized to their average value in the `asymptotic tail'
for  $I < I_c$.}
\end{figure}

Now let us discuss the behaviour of the fractal dimension $D_{-4}$ characterizing the
exploding strange attractors for gradually variing control parameter. This is shown in Fig.
\ref{fig_Dm4O1phi}  for oscillator O1 with parameter sets {\em (A)} and {\em (B)} for 
various choices of the phase shift $\Delta \phi$ of the periodic driving force for which
 the Poincar\'e sections were taken. Making the comparison more straightforward, the values $D_{-4} (I,\Delta \phi)$ for each $\Delta \phi$  are normalized to
 their average  values  taken in interval $I<I_c$, i.e.,  for the unexploded attractor. It has been
observed that for both discussed cases of the explosion the phase shift with $\pi$ produces
an almost rigid rotation of the Poincar\'e map of the strange attractor with the angle $\pi$.
This results in the agreement of the $D_{-4}$ values within the error bars for 
$\Delta \phi=0$ and $\pi$ 
and for $\Delta \phi=\pi/2$ and $3\pi/2$ that is illustrated  by a few
 points  for parameter set {\em (A)} in Fig.  \ref{fig_Dm4O1phi}.  In order to reduce the
CPU time the detailed analysis  has been performed on the base of $D_{-4}$ values obtained 
for the choices $\Delta \phi=0$ and $\pi/2$. In Fig.  \ref{fig_Dm4O1phi} one can see that 
there occurs a significant jump of the dimension $D_{-4}$ at $c=0$ when the explosion starts,
 i.e., when the approximately linear rise of the average maximal Lyapunov exponent begins
 in Figs. \ref{fig_lambO1A} and \ref{fig_lambO1B}. Furthermore, it is also seen in Fig.
 \ref{fig_Dm4O1phi} that the height of the jump is rather sensitive to the choice of the
phase shift $\Delta \phi$. Nevertheless, taking the weighted average   ${\bar D}_{-4}$
of the $D_{-4}$ values over the various phase shifts, the jump, i.e., the effect of
the start of the explosion remains still significant in both cases. One would need much
 better statistics and smaller error bars, i.e., orders of magnitude more vectors in the 
embedding space in order to decide  whether the singular
 behaviour of ${\bar D}_{-4}(c)$ is the jump at $c=0$ of an approximately steplike
function or that of a peaked function with a wider tail for $c>0$.
Anyhow, the effect, i.e., the jump of the average  dimension ${\bar D}_{-4}$ at 
the start of the explosion is significant even for the accuracy of our method.
One should also notice that the jump is of the factor of $\sim 3$ for the case {\em (A)}
and $\sim 6$ for case {\em (B)}, i.e., it becomes more expressed when the explosion 
of the strange attractor occurs for larger amplitude of the periodic driving force.
\begin{figure}[htb]
\epsfig{file=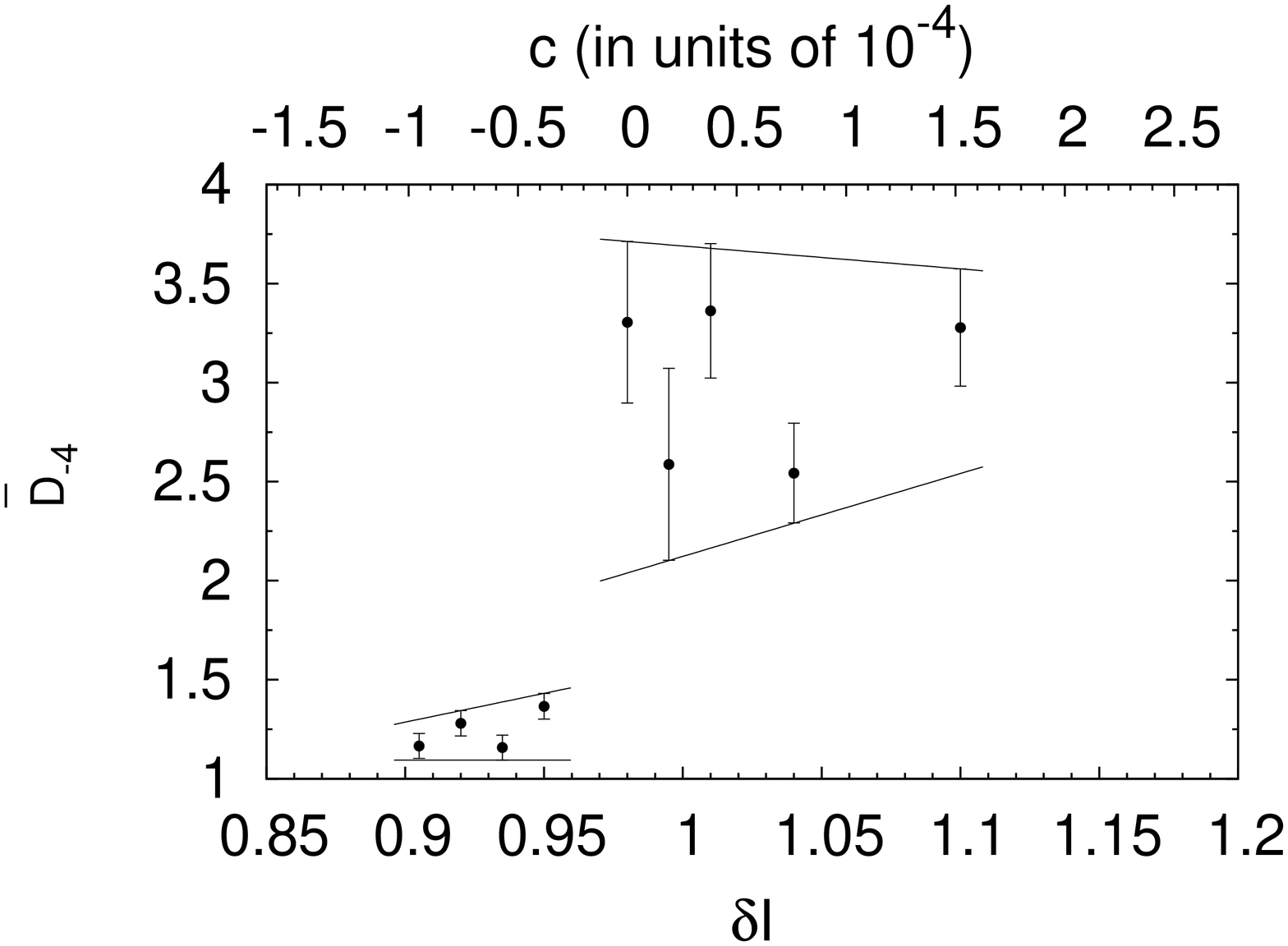,width=6.0cm,angle=-90}
\epsfig{file=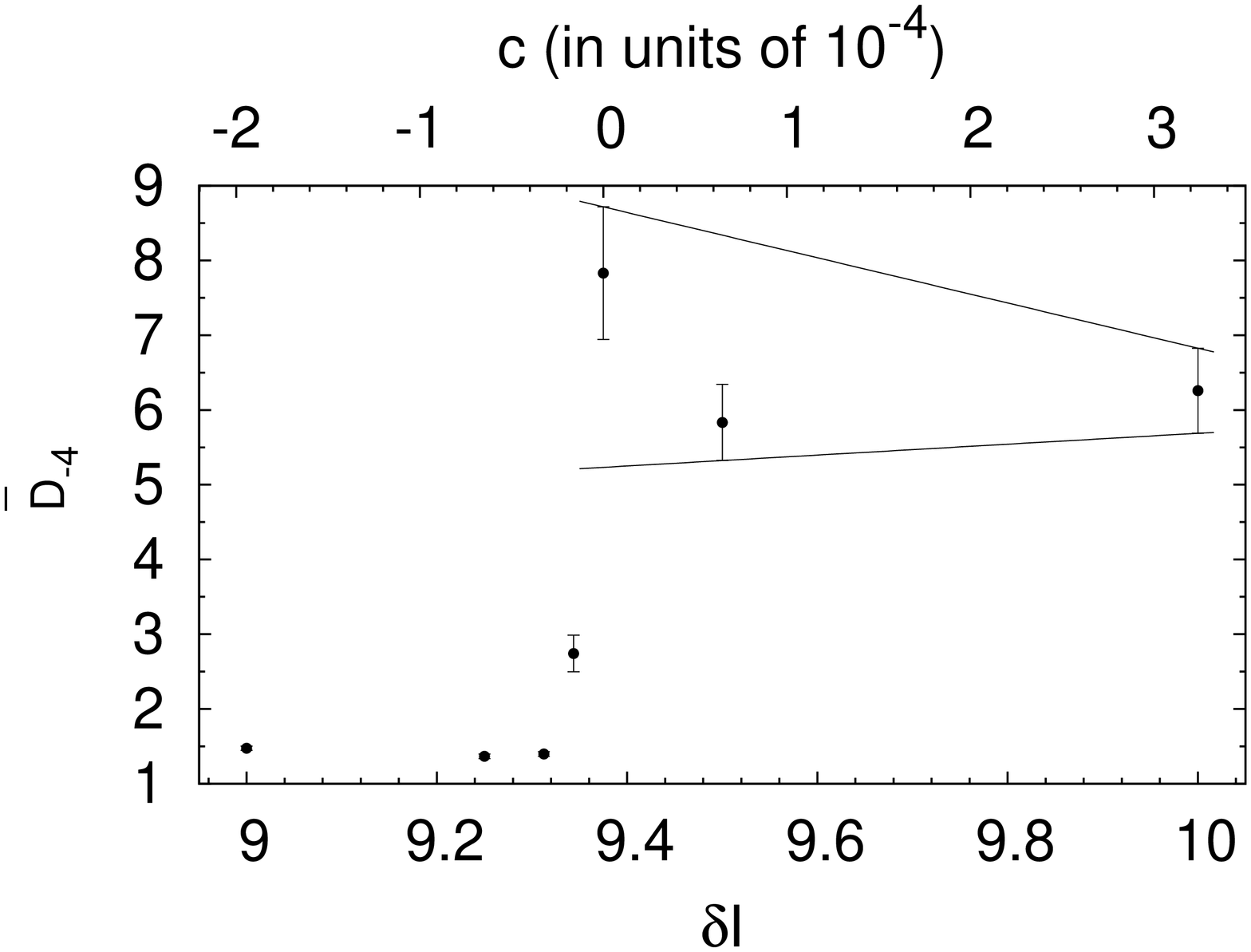,width=6.0cm,angle=-90}
\caption{\label{fig_Dm4O1av} The average generalized dimension ${\bar D}_{-4}$  vs. the 
shifted control parameter $\delta I$ (as given in Figs. \ref{fig_lambO1A} and \ref{fig_lambO1B}, respectively)  for the strange attractor belonging to oscillator O1 with parameter sets
{\em (A)} (top) and {\em (B)} (bottom).
 Also the scale $c=(I-I_c)/I_c$ is shown. The black lines are only to guide the eyes.
}
\end{figure}
The interval of almost linear increase of the maximal Lyapunov exponent
is of the width $\Delta c\approx  0.0004$ and $ 0.002$ for cases {\em (A)}
and {\em (B)}, respectively, i.e., the process of the explosion accomplishes during a rather
small relative change of the control parameter. The interval in which the singular behaviour
of ${\bar D}_{-4}$ occurs is even much  shorter, $(\Delta c)_{sing} \stackrel{<}{\sim} 0.0001$ for
 both cases. Even if  a few trajectory points burst out into the originally empty phase-space 
region, the value of $D_{-4}$ jumps suddenly.

\begin{figure}[ht]
\epsfig{file=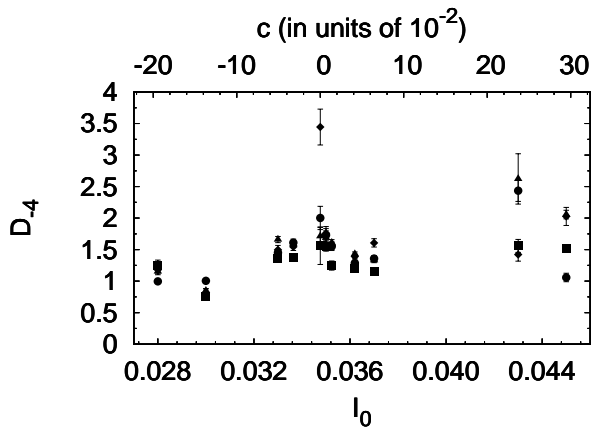,width=6.0cm,angle=-90}
\caption{\label{fig_Dm4O2phi} The generalized dimension $D_{-4}$ vs. the control
parameter $I_0$ for the strange attractor belonging to oscillator O2 with parameter set
{\em (C)} for various phase shifts: $\Delta \phi=0$ (circle),
$\pi/2$ (square), $\pi$ (triangle), and $3\pi/2$ (diamond). Also the scale $c=(I_0-I_{0c})/I_{0c}$ is shown.
 The $D_{-4}$ values are normalized to their average value in the `asymptotic tail'
for  $I_0 < I_{0c}$.}
\end{figure}
\begin{figure}[htb]
\epsfig{file=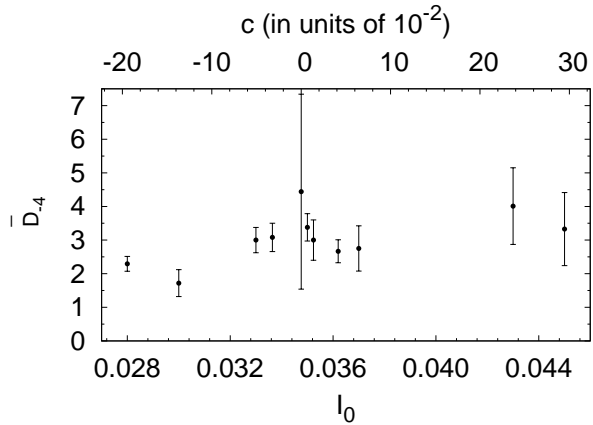,width=6.0cm,angle=-90}
\caption{\label{fig_Dm4O2av} The average generalized dimension ${\bar D}_{-4}$  vs. the 
control parameter $I_0$ for the strange attractor belonging to oscillator O2 with parameter set
{\em (C)}.
 Also the scale $c=(I_0-I_{0c})/I_{0c}$ is shown.
}
\end{figure}
For the strange attractor belonging to oscillator O2 with parameter set {\em (C)} we have also
determined the dependence of the dimensions $D_{-4}$ on the control parameter $I_0$ 
for various choices of the phase shift of the periodic external force
 (see Fig. \ref{fig_Dm4O2phi}). Here the Poincar\'e maps
belonging to various choices of the phase shift cannot be obtained from each other by a rigid
 rotation. This is reflected by the variation  of the $D_{-4}$ values with the phase shift
 $\Delta \phi$. Let us restrict our discussion to the interval $I_0\in\lbrack 0.033 , 0.038\rbrack $ surrounded by the wide regular 
windows from both sides. The $D_{-4}(I_0)$ values are more or less peaked at 
  $I_{0c}\approx 0.03476$ for which we have found the smallest
positive value of the average maximal Lyapunov exponent, although that peak is almost within
the estimated errors except the single point for the phase shift $\Delta \phi=3\pi/2$.
 A thoroughful looking through  the numerically evaluated scalings of 
$\ln C_{-4}$ vs. $\ln (R/R_{max})$ shows that in all cases only a single scaling interval exists.
The average ${\bar D}_{-4}$ values remain  constant within the estimated errors in  the interval $I_0\in\lbrack 0.033 , 0.038\rbrack $,  as shown in Fig. \ref{fig_Dm4O2av}.  This may be a hint
 to suggest that the effect if it is present at all
is so weak here that it cannot be seen by the  accuracy of our computations. Let us remind the
 reader that we have established for oscillator $O1$ that a decreasing amplitude of the control
 parameter causes a smaller jump of $D_{-4}$ when the explosion starts. 
For the strange attractor belonging to oscillator O2 under discussion  the control parameter
has extremely small values. This together with the
 fact that the highly populated bunches of the strange attractor are not separated for $I_{0c}$
with a really empty phase-space region can result in washing out the singularity
in the control-parameter dependence of ${\bar D}_{-4}$ within our computational  accuracy.

For the strange attractor with  parameter set {\em (C)} for oscillator O2
the widths of the intervals in which the average maximal Lyapunov exponent increases 
linearly are $\Delta c\approx 0.02$  and $\Delta c\approx 0.05$, respectively, to the left and to the right of the critical value $c=0$. This means that the explosion in this case is much
 less sudden than in the above  discussed cases for oscillator O1. The overall
 increase of the maximal Lyapunov exponent is a factor of $\sim 3$ and $\sim 2$ on the right-hand and left-hand sides of $I_{0c}$, respectively. This is quite similar to the factors of 
$\sim 2$ and $\sim 2.5$ of overall increase during the linear raise for the strange attractors
 with parameter sets {\em (A)} and {\em (B)}, respectively, for oscillator O1.
Afterall one has to conclude that the explosion of the strange attractor in the cases found by us
for oscillator O1 is much more rapid and violent with the change of the control parameter, then
the explosion for oscillator O2 with parameter set {\em (C)}.

\section{Summary}\label{conclu}

We have presented examples on a very rapid and violent explosion of  the strange attractor of a
 one-dimensional  externally driven damped anharmonic oscillator  when the  control parameter of the
explosion process, the amplitude of the 
strongly nonperturbative periodic driving force gradually increases by a relatively small amount.
As compared to its use in \cite{Ueda80}, the term `explosion' of the strange attractor is used
 by us in a rather phenomenologic and more general sense, disregarding of the  dynamical origin
 of the explosion. It is shown that the explosion process can be reliably characterized  by
the dependence on the control parameter of such phenomenologic characteristics as 
  the average maximal Lyapunov exponent ${\bar \lambda}$ and  the average  generalized dimension
 $\bar{D}_{-4}$. The former has been determined by  Benettin's method, the latter  by means of the 
combination of the embedding technique and the sandbox method. For comparison the exploding strange 
attractor discussed in \cite{Ueda80} has also been analysed in the same manner.

It has been shown that
the explosion of the strange attractor is accompanied by an approximately linear increase of the 
average maximal Lyapunov exponent ${\bar \lambda}$ in the cases presented by us as well as in the
 case given in \cite{Ueda80}. This reflects the increasing chaoticity of the
 strange attractor when it gradually builds up during the explosion process. The overall
 increment of the maximal Lyapunov exponent is of
the factor of cca. 2  to 3 in the various cases. In the cases presented by us the explosion 
is accomplished rather rapidly, after $\le 0.3$ per cent of  the relative change of the control
 parameter. As opposed to this,  in the case discussed in
  \cite{Ueda80} the explosion accomplishes much slowly, it needs   1 to 5 per cents of relative
 change of the control parameter.

  In the examples presented by us
also a rather sudden jump of the generalized dimensions $D_{q}$ with negative parameter $q$, 
in particular that of the average dimension ${\bar D}_{-4}$ occurs when the explosion sets on, while
 we have not seen such a singularity in the case given in \cite{Ueda80} within our computational
 accuracy. This disagreement may
 be explained, on the one hand, by the fact that in our examples the explosion starts from an attractor consisting 
of disjoint bunches which are separated by an  empty phase-space region, while the latter is
 only underpopulated but not empty just before  the  strange attractor discussed in  \cite{Ueda80}
 bursts out and therefore one expects a much smaller effect in the latter case. On the other hand, 
our computations have been performed by using a relatively small number of  $25000$ embedded vectors,
so that the weaker effect may not exceed the numerical errors.  
 The empty  phase-space region induces a gap in the string of the one-dimensional data
 for the determination of the generalized correlation sum $C_{-4}(R)$ and that results in the
 dominance of the short-distance scaling of  $\ln C_{-4}(R)$ for the yet unexploded strange 
attractor. When a few trajectory points start to occupy the originally empty phase-space region
the contributions of inter-bunch distances become overemphasized in the generalized correlation sum 
$C_{-4}$ because of the much lower occupation probabilities in the phase-space region between the 
bunches than in the bunches themselves and this results in a sudden jump of the average generalized
 dimension ${\bar D}_{-4}$. The larger is the control parameter,  the amplitude of the periodic
 driving force, the greater is the factor by which the dimension $\bar{D}_{-4}$ increases.
The smallness of the driving force in the case discussed in  \cite{Ueda80} may be an additional source
of such a weak effect of the explosion on ${\bar D}_{-4}$ that does not reveals itself under 
 the accuracy of our approach.

\section*{Acknowledgement}
The authors are   grateful       to S. Nagy for his valuable remarks.

\end{document}